\documentclass[
prd,aps,amsmath,superscriptaddress,nofootinbib,
]{revtex4}

\usepackage[hypertex]{hyperref}
\usepackage{graphicx}
\usepackage{epstopdf}
\usepackage{bm}
\usepackage{epsfig}
\usepackage{xspace}
\usepackage{color}
  \newcount\hour \newcount\minute
  \hour=\time \divide \hour by 60
  \minute=\time
  \newcommand{\mydate}{\ \today \ - \number\hour :\ifnum \minute<10 0\fi
\number\minute}
\newcommand{\nn}{\nonumber}
\newcommand{\beqn}{\begin{eqnarray}}
\newcommand{\eeqn}{\end{eqnarray}}
\newcommand{\be}{\begin{equation}}
\newcommand{\ee}{\end{equation}}

\newcommand{\ba}{\begin{array}{c}}
\newcommand{\bat}{\begin{array}{cc}}
\newcommand{\ea}{\end{array}}

\newcommand{\bi}{\begin{itemize}}
\newcommand{\ei}{\end{itemize}}

\newcommand{\Frac}[2]{\frac{\displaystyle #1}{\displaystyle #2}}

\newcommand{\cO}{{\cal O}}

\newcommand{\mF}{\mathcal{F}}

\newcommand{\mL}{\mathcal{L}}
\newcommand{\mM}{\mathcal{M}}
\newcommand{\mO}{\mathcal{O}}

\newcommand{\we}{\widetilde{e}}

\newcommand{\lsim}{\stackrel{<}{_\sim}}

\newcommand{\bea}{\begin{eqnarray}}
\newcommand{\eea}{\end{eqnarray}}
\newcommand{\bqa}{\begin{eqnarray}}
\newcommand{\eqa}{\end{eqnarray}}
\newcommand{\beq}{\begin{equation}}
\newcommand{\eeq}{\end{equation}}

\newcommand{\bear}{\begin{eqnarray}}
\newcommand{\eear}{\end{eqnarray}}
\newcommand{\ket}{\,\rangle}
\newcommand{\bra}{\langle \,}

\begin{document}
\setlength\baselineskip{12pt}

%%%%%%%%%%%%%%%%%%%%%%%%%%%%%%%%%%%%%%%%%%
%Define Title, Author, Address, Preprint#

%\preprint{
% \vbox{
% \hbox{MPP-2008-83}
% \hbox{arXiv:0807.4173         } }}

\title{
$\pi\pi$--scattering lengths at $\cO(p^6)$ revisited
\\[3mm]\mbox{}}

\vspace*{1cm}

\author{
Zhi-Hui Guo\footnote{Electronic address: zhguo@ifae.es}    }
\affiliation{
 Grup de F\'isica Te\`orica and IFAE,
Universitat Aut\`onoma de Barcelona, 08193 Barcelona, Spain\\
and Department of Physics, Peking University,Beijing 100871,
P.~R.~China\\[5mm]}

\author{
 Juan Jos\'e Sanz-Cillero\footnote{Electronic address: cillero@ifae.es} }
\affiliation{Grup de F\'isica Te\`orica and IFAE,
Universitat Aut\`onoma de Barcelona, 08193 Barcelona, Spain\\[5mm]}

%\date{\today\\ \vspace{1cm} }

%%%%%%%%%%%%%%%%%%%%%%%%%%%%%%%%%%%%%%%%%%
%Create the title page

\begin{abstract}

This article completes a former work where part of the $\cO(p^6)$
low-energy constants entering in the $\pi\pi$ scattering were estimated.
Some resonance contributions were missed in former calculations
and slight differences appeared with respect to our outcome.
Here, we provide the full results for all the contributing $\cO(p^6)$ couplings.
We also perform a reanalysis of the hadronic inputs used for the estimation
(resonance masses, widths...). Their reliability was checked
together with the impact of the input uncertainties on the
determinations of the chiral couplings and the scattering lengths $a^I_J$.
Our outcome is found in agreement with former works  though with slightly larger errors.
However, the effect in the final values of the $a^I_J$ is negligible
after combining them with the other uncertainties.
Based on this consistency, we conclude that
the previous scattering length determinations seem to be rather solid and reliable,
with the $\cO(p^6)$ low-energy constants quite under control. Nevertheless, the uncertainties found in the present work
point out the limitation on further improvements
unless the precision of the $\cO(p^6)$  couplings is properly increased.
\end{abstract}

%\pacs{
%11.15.Pg,
%%%Expansions for large numbers of components (e.g., 1/Nc expansions)
%12.39.Fe,
%%%Chiral Lagrangians
%11.55.Bq,
%%%Analytic properties of S matrix
%11.30.Rd,
%13.75.Lb
%}
%%
%% \keywords{$\pi\pi$ scattering, scattering length, chiral symmetry, large $N_C$}

\maketitle

%%%\newpage
%%%%%%%%%%%%%%%%%%%%%%%%%%%%%%%%%%%%%%%%%
\section{Introduction}

In a previous work~\cite{matching,matching-op6}, we provided a set of predictions
for some of the $\cO(p^6)$ low energy constants (LECs) $r_i$
related to the $\pi\pi$--scattering amplitude~\cite{bijri,bijff}.
The $\cO(p^6)$ chiral perturbation theory couplings  $r_2$, ... $r_6$  were determined there
in the large--$N_C$ limit~\cite{NC}
by means of once-subtracted partial-wave dispersion relations, where they were provided
in terms of the ratios of widths over masses,
$\Gamma_R/M_R^3$ and $\Gamma_R/M_R^5$.
The limit of a large number of colours $N_C\to\infty$~\cite{NC}, is a key ingredient
of the study, becoming the strong dynamics  greatly simplified and being the
dominant contribution provided by the tree-level meson exchanges.
At large $N_C$, the relevant resonances for $\pi\pi$--scattering are the I=1 vector
and the $\bar{u}u+\bar{d}d$ component of the scalar with I=0.
We will denote these large--$N_C$ states, respectively, as $\rho$ and  $\sigma$ all along the paper.
We will also consider just the contribution from the lightest multiplets:
the single resonance approximation will be assumed.

The new  predictions in Ref.~\cite{matching-op6} found that
some scalar  resonance contributions to the $\cO(p^6)$ LECs had been
actually missed in former estimates~\cite{bijri}.
This produced  small   variations on the
$\pi\pi$--scattering lengths~\cite{matching-op6},
of the order of the current errors~\cite{slgasser}. However, in order to
make a thorough analysis, we complete our former study and provide
the remaining $\cO(p^6)$ LECs contributing to $\pi\pi$--scattering.
In the standard  $\cO(p^6)$ chiral perturbation theory calculation~\cite{bijri},
the scattering lengths $a^I_J$  depend on $r_1$,... $r_6$.
Alternatively, if they are determined through the dispersive method
in Ref.~\cite{slgasser}, one needs to provide $r_{S_2}$ instead of $r_5$ and $r_6$.
%%%This LEC parametrizes a part of the scalar form factor at
%%%$\cO(p^6)$~\cite{slgasser,bijff}.

In addition, we have decided to redo the whole analysis of the resonance inputs.
This has enabled us with at least a minimal control of the uncertainties
in the LECs and   the corresponding scattering lengths.
This allows a better understanding of what are the relevant parameters that
determine the low energy scattering and the main sources of errors.
The vector mesons are found to fit very well within a $U(3)$ large--$N_C$ multiplet
and their properties are quite under control.
On the other hand, the current knowledge of the lightest large--$N_C$ scalars
is rather poor.  Thus, our revised  analysis of the $\pi\pi$--scattering lengths
is also motivated by the need of improving the current picture
of the lowest lying hadronic resonances~\cite{sigmaleut,sigmazheng,pdg}.

%%
%%Thus, the $\pi\pi$--scattering lengths play  an important  role
%%in the determination of the $\sigma$ pole position
%%through Roy equations~\cite{sigmaleut}.
%%
%%At large $N_C$, one finds that a resonance lagrangian
%%that couples scalar mesons to external scalar sources
%%(e.g, the operator $c_m\bra S \chi_+\ket $ in Ref.~\cite{ecker89})
%%leads unavoidably to the presence of scalar resonance
%%tadpoles~\cite{bernardtadp,juanjotadp}.
%%This effect is proportional to the quark masses and is absent
%%in the chiral limit. However, it is crucial for the
%%chiral corrections as one can see in the
%%large--$N_C$ determination of $F_\pi$ and $F_K$~\cite{juanjotadp}.
%%
%%An easy way to handle this issue is by performing a scalar meson field redefinition,
%%in order to remove the tadpole diagrams~\cite{juanjotadp}.
%%All the tadpole effects are then reabsorbed into shifts of the other
%%parameters of the mesonic lagrangian  (e.g., the pseudo-Goldstone
%%mass and wave function).
%%

In Section~II, we present the former LEC determinations and the
reanalysis that we propose.
The phenomenological inputs required for the determination
of the LECs are studied in full detail in Section~III,
keeping a careful control of the possible uncertainties.
Based on this,  the  values of the $\cO(p^6)$ low-energy constants $r_i$
are estimated  in Section~IV,  leading  to a series of new
predictions for the $\pi\pi$ scattering lengths in Section~V.
The various results and conclusions are  gathered in Section~VI.
Finally, some more technical details  are relegated to the Appendices.

\section{LEC estimates through resonance saturations}

The current calculations in chiral perturbation theory ($\chi$PT)
have already reached the two loop order~\cite{bijri,bijff}.  However,
in order to extract quantities such as the scattering lengths one needs
to eventually input the $\cO(p^6)$ low-energy constants.
Since in many cases it is not possible to determine these couplings
directly from the phenomenology,  one needs to extract their values
through alternative procedures.
One of the most usual ones is to consider large--$N_C$ estimates based
on phenomenological chiral lagrangians~\cite{bijri,ecker89,op6rxt}.
However, the control that one has on these lagrangians
and their large--$N_C$ estimation may be unclear.
Thus, the uncertainties of the observables under consideration
might be larger than what is actually quoted.

In this work, we propose the comparison of three sets of estimates of the
low-energy constants.  First we will review the currently used values
(set A)~\cite{bijri,bijff}
and then  we will present the new numbers after taking into account the scalar meson contributions
that were missing in former works (set B)~\cite{matching-op6}. However, for sake of consistency,
we will also redo the analysis of the hadronical inputs and provide the newly calculated LECs
(set C).
\begin{itemize}

\item{} \underline{\bf Set A:}

This is the group of estimates commonly employed in nowadays
calculations~\cite{bijri,bijff}.
The $\chi$PT couplings are assumed to be
determined by the resonance exchanges provided by the phenomenological lagrangian
\bear
\mL &=& \Frac{F^2}{4}\bra u_\mu u^\mu +\chi_+\ket
\nn\\
&& \, -\, \Frac{1}{4}\bra \hat{V}_{\mu\nu}\hat{V}^{\mu\nu}\ket
+\Frac{1}{2} M_V^{  2} \bra \hat{V}_\mu\hat{V}^\mu\ket \,\,
-\Frac{i g_V}{2\sqrt{2}}\bra \hat{V}_{\mu\nu}[u^\mu,u^\nu]\ket
+f_\chi \bra \hat{V}_\mu [u^\mu,\chi_-]\ket
\nn\\
&&   +\Frac{1}{2}\bra \nabla^\mu S \nabla_\mu S\ket
-\Frac{1}{2}  M_S^{ 2} \bra SS\ket
+c_d \bra S u_\mu u^\mu \ket +c_m \bra S \chi_+\ket \, ,
\label{lagr}
\eear
where $\bra ...\ket$ stands for trace in flavour space, $S$ and $\hat{V}^\mu$
account  respectively for the scalar and vector multiplets. The tensor
$u^\mu$ contains the chiral pseudo-Goldstone and $\chi_\pm$
is, in addition, proportional to the light quark masses. Their precise definitions
can be found in Refs.~\cite{bijri,bijff,ecker89}.
Notice that since the lagrangian does not contain resonance mass-splitting terms,
the  masses $M_R$ coincide with their chiral limit values,
which we denote as $\overline{M}_R$.  Nevertheless, it can be
straightforwardly extended to include the splitting due to the quark masses. This will be studied with full
generality in the next sections. From the comparison of the decays
$\rho\to\pi\pi$, $K^* \to K\pi$ and other processes,
Ref.~\cite{bijri} provided the set of parameters
\bear
M_V=770\, \mbox{MeV}\, , \qquad
g_V=0.09\, , \qquad f_\chi=-0.03  \, ,
\nn\\
M_S=983\, \mbox{MeV}\, , \qquad c_m=42\, \mbox{MeV}\, ,\qquad
c_d=32\, \mbox{MeV}\, .
\label{setAinputs}
\eear
Taking these inputs and the phenomenological lagrangian~(\ref{lagr}),
they produced the LEC estimates~\cite{bijri},
\bear
r_1^A=-0.6 \times 10^{-4}\, , \qquad \qquad&
r_2^A=1.3 \times 10^{-4}\, , \qquad \qquad
&r_3^A=-1.7 \times 10^{-4}\, ,
\nn \\
r_4^A=-1.0 \times 10^{-4}\, , \qquad \qquad&
r_5^A=1.1 \times 10^{-4}\, , \qquad \qquad
&r_6^A=0.3 \times 10^{-4}\, ,
\nn\\
\label{eq.setA}
\eear
where the authors already accounted in these numbers
the contribution to the $SU(2)$ LECs coming from the kaon and eta loops~\cite{bijri}:
\bear
r_1^K=\Frac{31 F_\pi^2}{5760 \pi^2 m_K^2}\, , \qquad\qquad
r_2^K=-\Frac{11 F_\pi^2}{2304 \pi^2 m_K^2}\, , \qquad\qquad
r_3^K=-\Frac{29 F_\pi^2}{7680 \pi^2 m_K^2}\, ,
\nn\\
r_4^K=-\Frac{F_\pi^2}{2560 \pi^2 m_K^2}\, , \qquad\qquad
r_5^K=\Frac{23 F_\pi^2}{15360 \pi^2 m_K^2}\, , \qquad\qquad
r_6^K=\Frac{  F_\pi^2}{15360 \pi^2 m_K^2}\, .
\nn\\
\label{SU2kaons}
\eear
These  will be also included in our analysis in order to have a clearer comparison
with the outcomes obtained in Refs.~\cite{bijri,slgasser}.

In addition to this, the dispersive approach used in Ref.~\cite{slgasser}
also  required the $\pi\pi$--scalar form-factor coupling
$r_{S_2}$, which was estimated to be~\cite{bijff}
\be
\label{setA2}
r_{S_2}^A\, =\,  -0.3 \times 10^{-4}\, ,
\ee
with a negligible contribution
$r_{S_2}^K= \Frac{F_\pi^2}{1152\pi^2 m_K^2}=0.03 \cdot 10^{-4}$ from
kaon and eta loops~\cite{bijff}.

\item{} \underline{\bf Set B:}

However, the resonance estimate from Refs.~\cite{bijri,slgasser}
was incomplete and some relevant
contributions were actually missing in their calculation.
The presence of the operator $c_m\bra S \chi_+\ket$ in the scalar meson lagrangian
produces a tadpole term proportional to the quark mass~\cite{bernardtadp,juanjotadp}.
Although its effect vanishes in the chiral limit, its contribution
becomes relevant at subleading chiral orders.

The  $\cO(p^6)$ LECs are recalculated here in full detail,
keeping all the possible contribution.  No term of the corresponding chiral order under study
is neglected.   By means of the partial-wave dispersion relations proposed
in Refs.~\cite{matching,matching-op6},
most of the $\cO(p^6)$ LECs in the $\pi\pi$--scattering
($r_{2},\,...\, r_6$)  are now fixed
in terms of the ratios,
\begin{eqnarray}
%%%\label{alphagamma}
\frac{\Gamma_R}{M_R^3}&=&\frac{\overline{\Gamma}_R}{\overline{M}_R^{\,\, 3}} \left[1+
\alpha_R\frac{m_\pi^2}{\overline{M}_R^{\,\, 2}}
+\gamma_R\frac{m_\pi^4}{\overline{M}_R^{\,\, 4}}+\cO(m_\pi^6)\right]\, ,
\nn \\
\nn \\
\label{beta}
\frac{\Gamma_R}{M_R^5}&=&\frac{\overline{\Gamma}_R}{\overline{M}_R^{\,\, 5}} \left[1+
\beta_R\frac{m_\pi^2}{\overline{M}_R^{\,\, 2}}+\cO(m_\pi^4)\right] \, ,
\label{KSRFdefi}
\end{eqnarray}
where $\overline{M}_R$ and $\overline{\Gamma}_R $ stand for the chiral
limit of $M_R$ and $\Gamma_R$, respectively. The constants
$\alpha_R$, $\beta_R$, $\gamma_R$ are quark mass independent and
rule the $m_\pi$  corrections in the ratios.
The formal expressions
for $r_{2},...r_{6}$~\cite{matching-op6} are provided later in Sections~\ref{sec.r5r6}--\ref{sec.r2}.
The resonance lagrangian~(\ref{lagr}) allows us then to compute
the resonance masses and widths at large $N_C$ in terms of the resonance couplings.
Thus,  the LEC predictions can be utterly rewritten in terms of the latter.
Using exactly the same inputs (\ref{setAinputs}) of set A,
one  obtains  then slightly different values
\bear
r_2^B=18  \times 10^{-4}\, , \qquad \qquad
&
r_3^B=0.9  \times 10^{-4}\, , \qquad \qquad
&r_4^B=-1.9 \times 10^{-4}\, ,
\nn\\
\label{setB1}
\eear
remaining $r_5$ and $r_6$ unchanged, i.e., $r_5^B=r_5^A$ and $r_6^B=r_6^A$
as in Eq.~(\ref{eq.setA}).
Although the variation in $r_3$,... $r_6$ with respect to Set A
is relatively mild, $r_2$ changes by an order of magnitude.
This points out the need of a more detailed investigation of the
stability of the estimation under modifications in the inputs.
%%
%%This already included the kaon and eta loop corrections (\ref{SU2kaons}).

To complete the calculation we needed to extract $r_1$ (for the direct $\chi$PT
calculation~\cite{bijri})  and $r_{S_2}$ (for the dispersive method~\cite{slgasser}).
These two couplings were out of the reach of the partial-wave dispersion relations
proposed in our former works~\cite{matching,matching-op6}.
Here we have used the lagrangian~(\ref{lagr}). We have
performed the explicit field theory calculation of
the  $\pi\pi$ scattering $A(s,t,u)$ and the scalar form factor $\mF_S(s)$.
The details of the calculation are shown in Appendix~A.

The corrected determinations derived from the phenomenological
lagrangian~(\ref{lagr}) result
\bqa
\label{r1nosplitana}
r_1 &=&  -\Frac{4 g_V^2F^2}{M_V^2} \left(2+ \frac{4\sqrt{2}f_\chi}{g_V}\right)^2
\, - \,  \Frac{16 c_d c_m ( 8 c_d^2-17 c_d c_m + 12 c_m^2)}{M_S^4}  \,,
\\
%\eqa
%\bqa
\label{rs2nosplitana}
r_{S_2} &=& \Frac{ 8 c_m (c_m-c_d) F^2}{M_S^4}
\,-\, \Frac{32 c_d^2 c_m^2}{M_S^4}
\, ,
\eqa
which for the set A inputs in Eq.~(\ref{setAinputs}) yield
\bqa \label{r1nosplitnum}
r_1^B \,\,=\,\, -2.1 \times 10^{-4}\,,
\qquad\qquad
r_{S_2}^B \,\, =\,\,  -0.3 \times 10^{-4}
\, .
\eqa
%%We have actually found   a typo in the sign of the $r_{S2}$ determination
%%provided  at Ref.~\cite{bijff} and used later in Ref.~\cite{slgasser}.
Our prediction  $r_{S2}^B$
agrees accidently the numerical value $r_{S_2}^A$ reported in Ref.~\cite{bijff} and
the variation in $r_1$ is also found to be  small.

\item{} \underline{\bf Set C:}

The analysis can be further refined.  In addition to performing the full computation of the LECs (without neglecting any contribution),
a more careful examination of the
phenomenological inputs also seems convenient.
A more general description of the mesonic interactions is required.
For instance, the lagrangian
considered in Eq.~(\ref{lagr}) does not take into account the fact that
in physical QCD there is a mass splitting within the $U(3)$ resonance multiplet.
This quark mass effect utterly contributes to the
scattering lengths at the same order as the terms already included
in the determinations from sets A and B.
A similar thing happens with the resonance widths, which accept a more general
quark mass splitting pattern.

Actually, it is possible to describe most of the U(3)  breaking without
relying in any particular resonance lagrangian realization.
Nevertheless, without lost of generality, we will study the vector
resonances in both the Proca and antisymmetric formalisms~\cite{spin1fields}.
Although the two representations are physically equivalent, small discrepancies
may appear when considering $m_\pi^2$ corrections only up to a given order.
The slight difference that appears when rearranging the experimental
information from one formalism to the other will serve us as an estimate
of the residual error due to higher order $m_\pi^2$ contributions.
We will also perform a  general analysis of the
splitting in the resonance masses and widths. This will allow us to fix
their dominant chiral corrections, being reflected in a
more accurate estimate  of the LECs $r_3$,... $r_6$.
However, it will be impossible for us to have a full control of
the subdominant chiral corrections
--that are going to affect $r_1$, $r_2$ and $r_{S_2}$--.
Therefore, we will have to utterly rely
on estimates of these LECs based on phenomenological resonance lagrangians
like that in Eq.~(\ref{lagr}).
This Set C reanalysis of the resonance parameters will be shown in detail
in the next Section.

\end{itemize}

\section{Phenomenology of the resonance parameters}
\label{pheno}

%\subsection{Determination of the resonance parameter at LO and NLO}
%
%
%\subsection{Chiral corrections to the meson properties}

\subsection{Mass splitting up to $\cO(m_P^2)$}

In the large--$N_C$ limit, the mass splitting
of the resonance multiplets can be described at leading order by a single
operator $e_m^R$~\cite{mass-split}
\bqa
\label{eq.eRm}
 -\frac{\overline{M}_R^{\,\,2}}{2}\bra R R\ket \, + \, e_m^R \bra R R \chi_+\ket \,,
\eqa
which leads at large $N_C$ to the mass eigenstates
\bqa
M^2_{I=1} &=& \overline{M}_R^{\,\,2} - 4 e_m^R m_\pi^2\,\,\,
+\,\,\, \cO(m_P^4)   \,,
\nonumber\\
M^2_{I=\frac{1}{2}} &=& \overline{M}_R^{\,\,2} - 4 e_m^R m_K^2\,\,\,
+\,\,\,\cO(m_P^4)  \,,
\nonumber\\
M^{(\bar{s}s)\,\,\, 2}_{I=0} &=& \overline{M}_R^{\,\,2}
- 4 e_m^R \,\, (2 m_K^2-m_\pi^2)\,\,\,
+\,\,\,\cO(m_P^4)  \,.
\label{masssplit}
\eqa
and  $
M^{(\bar{u}u+\bar{d}d)}_{I=0} = M_{I=1}$.

In the case of the vector resonance multiplet, $e_m^V$
can be fixed through the physical $\rho(770)$ and
$K^*(892)$ masses, respectively
${  M_\rho= 775.5\pm 0.3  }$~MeV and ${  M_{K^*}=893.6\pm 1.9  }$~MeV~\cite{pdg}.
This provides the  estimates
\be
\label{eV}
\overline{M}_V \, \,= \,\,  764.9 \pm 0.5  \, \mbox{MeV}\, ,
\qquad
e_m^V\, \, = \,\,  -0.217 \pm 0.004 \,.
\ee
These values and  Eq.~(\ref{masssplit}) lead to a prediction
for the remaining two states, $M^{(\bar{s}s)}_{I=0}=998$~MeV and
$M^{(\bar{u}u+\bar{d}d)}_{I=0}=M_\rho=776$~MeV.
This is in pretty good agreement with the $\phi(1020)$ and $\omega(782)$ masses,
respectively $M_\phi=1019$~MeV and $M_\omega=783$~MeV.
Hence, alternatively, one can think of extracting the chiral limit resonance mass
and the splitting term $e^V_m$ by means of the $\rho(770)$ and
$\phi(1020)$ masses, with ${  M_\phi= 1019.455 \pm 0.020 }$~MeV~\cite{pdg}:
\bear
\overline{M}_V \, \,= \,\, 763.7 \pm 0.5 \, \mbox{MeV}\, ,
\qquad
e_m^V\, \, = \,\,  -0.2414 \pm 0.0022 \,,
\label{eVbis}
\eear
with the prediction $M_{I=1/2}=906$~MeV,  in relatively
good agreement with the experimental $K^*(892)$ mass~\cite{pdg}.

The physical masses are indeed slightly different to their large--$N_C$ values.
Thus, the relations in Eq.~(\ref{masssplit}) are not exactly fulfilled
for the experimental inputs as they gain subleading  contributions in $1/N_C$.
The combination of Eqs.~(\ref{eV}) and (\ref{eVbis})
provide our most reliable estimate of the large--$N_C$ parameters.
Considering an interval that covers both determination, we will take as inputs
from now on the values
\be
\label{eV2}
\overline{M}_V \, \,= \,\,  764.3 \pm 1.1  \, \mbox{MeV}\, ,
\qquad
e_m^V\, \, = \,\,  -0.228\pm 0.015 \,,
\ee
whose errors exceed those that come from purely the experimental uncertainties
in Eqs.~(\ref{eV}) and (\ref{eVbis}).

In the case of the scalars, it is relatively straightforward to
identify the lightest $I=1$ resonance with the $a_0(980)$:
${   M_{I=1}=  984.7\pm 1.2 }$~MeV~\cite{pdg}.
The remaining states of the multiplet are more cumbersome to pin down.
The scalar spectrum provided in the large--$N_C$ study of Ref.~\cite{mass-split}
leaves the light and broad scalar resonances $\sigma$ and $\kappa$
out of any classification multiplet. In order to avoid the problem
of the mixing of iso-singlet scalars, the analysis is performed with
the $I=1/2$ state, whose mass  in the large--$N_C$ limit is also poorly known.
The broad  $\kappa(800)$ seems to be a possible candidate although the first
clear $I=1/2$ scalar resonance signal is provided by the $K^*_0(1430)$~\cite{pdg}.
Hence, we take the conservative estimate
$M_{I=1/2}= 1050\pm 400$~MeV, which ranges from the $\kappa$
up to the $K^*_0(1430)$ mass.  This leads to the predictions
\bqa
\label{eS}
\overline{M}_S \,=\, 980 \pm 40 \, {\rm{MeV}}\,,\qquad e_m^S\,=\, -0.1 \pm 0.9 \,.
\eqa

\subsection{The splitting of the vector resonance  decay width up to $\cO(m_P^2)$ \label{avsection}}

The calculation of the quark mass corrections to the width
is slightly more complicate and one needs to make
some assumption on the structure of the interaction.

If we use the phenomenological lagrangian~(\ref{lagr}),
the vector decay width into two light pseudo-scalars,
$V \to \phi_1 \phi_2$,  shows the   structure
\bqa
\label{avdef}
\Gamma_{V\to\phi_1\phi_2} &=& C_{V 1 2}
\,\, \times\,\,
\Frac{ M_V^{\,\, n} \, \rho_{\rm V 12}^3 }{48\, \pi \,  F_{1}^2 \, F_{2}^2}
\,\,\, \lambda_{V\pi\pi}^2\,\,\, \left[1 + \epsilon_V \frac{m_1^2+m_2^2}{2 \overline{M}_V^2}
\,\,\,+\,\,\, \cO(m_P^4)\right]^2\,,
\eqa
with the phase-space factor
\be
\rho_{\rm V 12}= \sqrt{\bigg(1-\frac{(m_1+m_2)^2}{M_V^2}\bigg)
\bigg(1-\frac{(m_1-m_2)^2}{M_V^2}\bigg)}\,.
\ee
The $F_i$ are the physical decay constants  of the $\phi_i$
pseudo-Goldstones ($F_\pi\simeq 92.4$~MeV
and $F_K\simeq 113$~MeV)  and they appear due to the
large--$N_C$ wave function renormalization of the light
pseudo-scalars~\cite{matching-op6,juanjotadp}. $M_V$ and $m_i$ corresponds, respectively,
to the physical vector and pseudo-scalar masses.
$\lambda_{V\pi\pi}$ is the $V\phi_1\phi_2$ coupling that rules
the decay amplitude in the chiral limit
(e.g. $g_V$ in Eq.~(\ref{lagr})~\cite{bijri,spin1fields}) whereas $\epsilon_V$ rules
the quark mass corrections to the amplitude (provided for instance
by the coupling $f_\chi$ in Eq.~(\ref{lagr})~\cite{bijri}). $C_{V 1 2}$ is the corresponding
Clebsch-Gordan: $C_{\rho\pi\pi}=1$, $C_{{K^*} K\pi}=3/4$ and $C_{\phi K\overline{K}}=1$.
The vector mass scaling $n$ is an integer that
depends on the resonance lagrangian under consideration:
$n=5$ in the Proca formalism~\cite{bijri,spin1fields} and $n=3$ in the
antisymmetric tensor representation~\cite{ecker89,op6rxt}.
Finally, one power of $\rho_{V 1 2}$ comes from phase-space and the other
two powers are due to the $J=1$ spin polarization sum of the
squared modulus of the amplitude,
$\sum_{\epsilon}|\mM_{V\to 12}|^2
= \sum_\epsilon |2 \epsilon_\mu p_1^\mu \widetilde{\mM}|^2
= M_V^2 \rho_{V12}^2 |\widetilde{\mM}|^2$, where here the $\epsilon_\mu$
denote the vector meson polarizations.

Thus, in the Proca field realization from Eq.~(\ref{lagr})
one has~\cite{bijri,bijff,spin1fields}
\be
n\,=\,5\,, \qquad\qquad \qquad
\lambda_{V\pi\pi}\, =\, g_V\,,\qquad\qquad \qquad
\epsilon_V \, =\,   4\sqrt{2} f_\chi/g_V\, .
\ee
On the other hand, in the antisymmetric tensor representation employed
in Resonance Chiral Theory (R$\chi$T)~\cite{ecker89,op6rxt}
one has $\lambda_{V\pi\pi}=G_V$, $n=3$ and the quark mass corrections $\epsilon_V$
provided by a combination of some extra $V\phi\phi$ operators,
$i\lambda_8^V \bra V_{\mu\nu} \{ \chi_+ , u^\mu u^\nu\}\ket$,
$i \lambda_9^V \bra V_{\mu\nu} u^\mu\chi_+u^\nu  \ket$~\cite{op6rxt}.
The detailed discussion on the quark mass splitting operators
in the antisymmetric formalism is relegated to Appendix~\ref{lambdaV8}.
Abusing of the notation, we will denote $f_\chi$ as the  effective combination
of resonance chiral theory couplings such that for the antisymmetric formalism ($n=3$) we still keep
$\epsilon_V= 4\sqrt{2} f_\chi/\lambda_{V\pi\pi}= 4\sqrt{2} f_\chi/G_V$.

In order to remain as general as possible,
both vector formalism scalings, $n=5$ and $n=3$, are discussed.
The following inputs will be used all throughout this paper to estimate the
uncertainties of our results~\cite{pdg}:
\bear
F_\pi =    92.4 \pm 0.3  \, \mbox{MeV}\,,&
\qquad &F_K  =   113.0 \pm 1.0  \, \mbox{MeV}\,,
\nn\\
m_\pi =  137.3 \pm 2.3  \, \mbox{MeV}\,,&
&m_K=  495.6 \pm 2.0   \, \mbox{MeV}\,,
\nn\\
\Gamma_{\rho \to \pi\pi}=   149.4 \pm 1.0  \, \mbox{MeV}\,,&
\qquad &\Gamma_{K^*\to \pi K}=   50.6 \pm 0.9  \, \mbox{MeV}\, ,
\eear
together with the mass parameters $\overline{M}_V$ and $e_m^V$
derived in Eq.~(\ref{eV2}).

The combination of the experimental $K^*$ and $\rho$ widths
allows us to fix the parameters $\lambda_{V\pi\pi}$ and
$\epsilon_V$:~\footnote{
Indeed, the system of quadratic equations has
two possible solutions  for $\lambda$ and $\epsilon_V$.
In one case $\epsilon_V$ is much larger ($\epsilon_V\simeq -10$) than in the other.
Since this corresponds to huge chiral corrections,
we will only keep the small $\epsilon_V$ solution ($|\epsilon_V|\lsim 1$).}
\bear
\label{splitV1}
\lambda_{V\pi\pi}^{\rm n=5}=g_V=0.0846 \pm 0.0008\, ,& \qquad \quad
& \epsilon_V^{\rm{n=5}}=0.01 \pm 0.09\,, \qquad \mbox{for n=5}\,,
\\
\label{splitV2}
\lambda_{V\pi\pi}^{\rm n=3}=G_V=  63.9 \pm 0.6 \, \mbox{MeV} \, ,& \qquad \quad
&\epsilon_V^{\rm{n=3}} =0.82\pm 0.10 \, ,
\qquad \mbox{for n=3}\,.
\eear
For $n=5$, this corresponds to the coupling $f_{\chi}= (0.2 \pm 1.3) \times 10^{-3} $,
to be compared to the determination
$f_\chi = -0.03$ from Ref.~\cite{bijri}~\footnote{
If, alternatively, we consider the same inputs as Ref.~\cite{bijri}
together with $F_K=110 \rm{MeV}$, then one gets
$f_\chi/g_V \simeq -0.03$, leading  to $f_\chi \simeq -0.003$.}.
In the case of the antisymmetric formalism, one obtains
$f_\chi^{n=3}= 9.3\pm 1.1 $~MeV.

The only difference between the Proca   (n=5)
and the antisymmetric tensor formalism  (n=3)
is that  in the latter the $\rho\to\pi\pi$ width carries the factor
$G_V^2 \left[1+\epsilon_V^{n=3}\frac{m_\pi^2}{\overline{M}_V^{\,\,2}}\right]$.
Instead, in the Proca case, this is replaced by
$g_V^2 M_\rho^2
\left[1+\epsilon_V^{n=5}\frac{m_\pi^2}{\overline{M}_V^{\,\,2}}\right]
\simeq g_V^2 \overline{M}_V^{\,\,2}
\left[1+(\epsilon_V^{n=5}-4 e^V_m)\frac{m_\pi^2}{\overline{M}_V^{\,\,2}}\right]$.
Up to higher chiral order corrections, one finds a pretty good agreement between
both of them, getting identical results for
$G_V=63.9$~MeV and $g_V \overline{M}_V= 64.7$~MeV,
and for $\epsilon_V^{n=3}=0.8$ and $(\epsilon_V^{n=5}-4 e_m^V)=0.9$.

The obtained values of $\lambda_{V\pi\pi}$ and
$\epsilon_V$ lead us to a prediction for the $\phi \to K \overline {K}$
decay width:
\bqa
\Gamma_{\phi \to K \overline{K}}\,\,= \,\,4.1\pm 0.8 \, \mbox{MeV} \, ,
\eqa
for both Proca and antisymmetric formalisms.
This result is perfectly consistent with the experimental value
$\Gamma_{\phi \to K \overline{K}}^{\rm Exp} = 3.54\pm 0.10 $~MeV~\cite{pdg}.
At large--$N_C$, the $\phi(1020)$ is identified with the $\bar{s}s$  component of the $I=0$ vector.

Likewise, the determination of the coupling $\lambda_{V\pi\pi}$
produces  for the $\rho\to\pi\pi$ decay width  the chiral limit prediction,
\bear
\overline{\Gamma}_\rho =
\Frac{ \lambda_{V\pi\pi}^2 \, \overline{M}_V^{\,\, n}}{ 48 \pi F^4}
&\longrightarrow\qquad &
\overline{\Gamma}_\rho=  181.9 \pm 3.0 \,  {\rm{MeV}} \, , \qquad \mbox{(n=5),}
\\
&&\overline{\Gamma}_\rho=  177.8 \pm 2.5 \, {\rm{MeV}}\, , \qquad \mbox{(n=3),}
\eear
where we have used the value of $F=   90.8 \pm 0.3$ MeV derived in Sec.~\ref{sec.Fpi}.

\subsection{The decay width for the scalar resonance}

In the case of scalar resonance decays,
the constraints to the structure of the width
are given just by an overall phase-space factor  $\rho_{V12}$
and the $F_i^{-1}$ factors due to the  $\phi_i$ wave--function renormalizations of the
pseudo-Goldstones~\cite{juanjotadp}.

The issue that arises here is the current poor knowledge on the structure
of the scalar sector in the large--$N_C$ limit.
One needs then to be assisted by some phenomenological lagrangian.
In the present work, we will assume that our scalar interactions
are ruled by the action from Eq.~(\ref{lagr}) together with  the mass
splitting operators $e^S_m$ from Eq.~(\ref{eq.eRm}).  Although many works have tried to pin down
the values of the two scalar couplings, $c_d$ and $c_m$,
in Eq.~(\ref{lagr})~\cite{ecker89,cdcm,cdcm2,ND,cd-Kaiser,Sdecays},
still there are no widely accepted results.

The $a_0 \to \pi^0 \eta$ decay seems to be
the only reliable process to estimate $c_d$ and $c_m$.
However, this only fixes a combination of the two
and one needs to add extra theoretical information.
Thus, the study of the high-energy behaviour of the $K\pi$,
$K\eta$, $K\eta'$ scalar form-factors performed in Ref~\cite{cdcm} provided the
constraints $c_m=c_d$ and $4 c_d c_m=F^2$.
However,  the first constraint seems {\it  a priori} less reliable, as the
analysis did not include all the possible quark mass
operators contributing at that order (these extra terms
in the lagrangian were derived later in Ref.~\cite{op6rxt}). On the other hand, the
relation $4 c_d c_m=F^2$ is more solid, as it stems from the
high-energy analysis of the scalar form-factor  in the chiral limit.

Hence,  the constraint $4 c_d c_m=F^2$ and the
$a_0 \to \pi \eta$ decay width  are used to fix the values of $c_d$ and $c_m$.
Thus, the lagrangian~(\ref{lagr}) yields
%%derived  from the lagrangian in Eq.(\ref{lagr}) is
%%found to be
%%\bqa
%%\Gamma_{a_0 \to \pi \eta}=  (\cos\theta-\sqrt2\sin\theta)^2
%% \, \times\,
%%\Frac{c_d^2\, M_a^3\, \rho_{a_0\pi\eta}}{24\pi F_\pi^2\, F_\eta^2}
%%\,\,\left( 1+ \frac{(2\frac{c_m}{c_d}-1) m_\pi^2-m_\eta^2}{M_a^2} \right)^2
%%,
%\eqa
\bqa
\Gamma_{a_0 \to \pi \eta}=
%%
%%(\cos\theta-\sqrt2\sin\theta)^2
%%
\bigg(\Frac{F_1\cos\theta_1-\sqrt2F_8\sin\theta_8}{F_1F_8\cos{(\theta_1-\theta_8)}}\bigg)^2
 \, \times\,
\Frac{c_d^2\, M_{a_0}^3\, \rho_{a_0\pi\eta}}{24\pi F_\pi^2}
\,\,\left( 1- \frac{m_\pi^2+m_\eta^2}{M_{a_0}^2}
+ (\epsilon_S+2) \frac{ m_\pi^2}{M_{a_0}^2} \right)^2
, \nn\\
\label{eq.a0-width}
\eqa
where the lagrangian from Eq.~(\ref{lagr}) yields
$\epsilon_S+2 =2 c_m/c_d$.
The $\eta$--$\eta'$ mixing is given in the basis of
the octet $\eta_8$ and  singlet $\eta_1$ by~\cite{fetaleut,thetaeta,thetaeta2}:
\bqa
\bigg(\begin{array}{c}
  \eta \\
 \eta'
\end{array} \bigg) = \frac{1}{F}\bigg(\begin{array}{cc}
                       F_8 \cos\theta_8 & -F_1 \sin\theta_1  \\
                       F_8 \sin\theta_8 & F_1 \cos\theta_1
                     \end{array} \bigg)
                     \bigg(\begin{array}{c}
                             \eta_8 \\
                             \eta_1
                           \end{array}
                      \bigg)\,.
\eqa
%\bqa
%%%
%%%\eta=-\sin\theta\,\eta_1 + \cos\theta\,\eta_8\,,\qquad \eta'=\cos\theta\,\eta_1 + %%\sin\theta\,\eta_8\,.
%%%
%\eta=-\Frac{F_1}{F}\, \sin\theta_1\,\eta_1 + \Frac{F_8}{F}\, \cos\theta_8\,\eta_8\,,\qquad \eta'=\Frac{F_1}{F}\, \cos\theta_1\,\eta_1 + \Frac{F_8}{F}\, \sin\theta_8\,\eta_8\,.
%\eqa
We consider the inputs
$F_1=(1.1\pm 0.1)\, F_\pi$, $\theta_1=(-5\pm 1)^\circ $,
$F_8=(1.3\pm 0.1)\, F_\pi$, $\theta_8=(-20\pm 2)^\circ$~\cite{fetaleut,thetaeta,thetaeta2},
${   \Gamma_{a_0 \to \pi \eta} =  75 \pm 25  }$~MeV,
$m_\eta= 547.9$~MeV~\cite{pdg},  together with the inputs considered
in previous sections. Most of  the error in the next determinations comes from our
poor knowledge on the $a_0(980)$ decay width.

Relying on the estimate of the $a_0\to\eta \pi^0$ decay width from the
phenomenological scalar lagrangian~(\ref{lagr})~\cite{bijri,ecker89} and the
theoretical constraint $4c_d c_m= F^2$~\cite{cdcm,cdcm2,PI:02}, one gets the values
\bqa
c_d\, =\,   26 \pm 7  \, {\rm{MeV}}\,,\qquad c_m\, =\, 80 \pm 21 \,{\rm{MeV}}\,.
\eqa
As the error is dominated by the large $a_0(980)$ width uncertainty,
we find that the determination of $c_d$ and $c_m$ is completely unaffected
by whether one considers the constraint $4 c_d c_m=F^2$ or
the approximation $4 c_d c_m=F_\pi^2$.
The chiral limit $F$ of the pion decay constant $F_\pi$  is computed
in the next section.  Indeed, the determination of $c_d$ is rather
model independent since the $(\epsilon_S+2)$ term in Eq.~(\ref{eq.a0-width})
has little impact in the $a_0$ decay width.  Thus,
if the $(\epsilon_S+2)$ term is neglected one gets $c_d=31$~MeV.

Taking this into account, the  $\sigma \to \pi \pi$  decay width is given
at large $N_C$ by
\be
\Gamma_\sigma \, = \, \Frac{3 \, \rho_{\sigma\pi\pi}\, c_d^2\,  M_\sigma^3\,
}{16\pi F_\pi^4}\,\left(1+\epsilon_S\frac{m_\pi^2}{\overline{M}_S^{\,\, 2}}
\right)^2\, ,
\ee
where by $\sigma$ we denote the iso-singlet $(\bar{u} u+\bar{d}d)$ scalar,
without strange quark
content. The quark mass  correction $\epsilon_S$ and the chiral limit of the width
are then provided by
\bqa
\label{splitS}
\epsilon_S=2\left(\frac{c_m}{c_d}-1\right) = 4 \pm 3\, ,
\qquad
\overline{\Gamma}_\sigma \, = \,
\Frac{3 \, c_d^2\,  \overline{M}_S^{\,\, 3}}{16\pi F^4}
\,\,  = 600 \pm 300 \,\mbox{MeV. }
\eqa

\subsection{Chiral corrections to $F_\pi$}
\label{sec.Fpi}

In order to extract the LECs related with $m_\pi^2$ corrections,
we will need to know the quark mass dependence of $F_\pi$,
which can be parametrized in the general form
\be
\label{Fpi-F}
F_\pi\,\,\, =\,\,\, F\,\, \left[ \, 1
\,\, +\,\, \delta F_{(2)}\, \Frac{m_\pi^2}{\overline{M}_S^{\,\, 2}}
\,\, +\,\, \delta F_{(4)}\, \Frac{m_\pi^4}{\overline{M}_S^{\,\, 4}}
\,\, +\,\, \cO(m_\pi^6)\, \right]\, .
\ee
The pion decay constant $F_\pi$ appears in the calculation
when one takes into account the pion wave-function renormalization  $Z_\pi$
that occurs at large $N_C$, which
obeys $F_\pi = F \cdot Z_\pi^{-1/2}$~\cite{matching-op6,juanjotadp}.

The scalar lagrangian in Eq.~(\ref{lagr})~\cite{ecker89} and
the mass splitting from Eq.(\ref{masssplit}) yield
\be
\delta F_{(2)}=\Frac{4 c_d c_m}{F^2}\, ,
\qquad \qquad
\delta F_{(4)}= \Frac{8 c_d c_m}{F^2}\left(
\Frac{3 c_d c_m}{F^2}-\Frac{4 c_m^2}{F^2}\right)\,
+  \Frac{16 c_d c_m e_m^S}{F^2} \, .
\label{eq.Fpi-ecker}
\ee
Although  $\delta F_{(2)}$ has the most general structure,
this is not true for $\delta F_{(4)}$.
If a more general set of scalar
operators $\lambda^S\mO_S$, $\lambda^{SS}\mO_{SS}$,
$\lambda^{SSS}\mO_{SSS}$  were allowed in the lagrangian~\cite{op6rxt},
$\delta F_{(4)}$  would gain a whole series of new contributions.

Substituting our former determinations of $c_d$, $c_m=F^2/4 c_d$ and $e_m^S$
in Eq.~(\ref{eq.Fpi-ecker}), one gets
\be
\label{dF2+dF4}
\delta F_{(2)}=1\, ,
\qquad \qquad
\delta F_{(4)}= -5\pm 5 \, ,
\ee
where the large uncertainty comes both from $e_m^S$ and the 25\%
error in $c_d$. For $\delta F_{(4)}$ one can use indistinctly
$c_m= F^2/4 c_d$ or $c_m=F_\pi^2/4 c_d$,
as the difference results negligible and goes to the next order in
the $m_\pi^2$ expansion. By means of Eqs.~(\ref{Fpi-F}) and (\ref{dF2+dF4})
it is then possible to recover the large--$N_C$ value for
the pion decay constant in the chiral limit, $F=90.8\pm 0.3$~MeV, in
agreement with former large--$N_C$ $\chi$PT determination~\cite{fetaleut}.

\subsection{NLO chiral symmetry breaking parameters $\alpha_R$, $\beta_R$}

Combining the information obtained from
the mass and width splittings, one can now extract
the corresponding quark mass corrections in the ratios $\Gamma_R/M_R^3$ and $\Gamma_R/M_R^5$
defined in Eq.~(\ref{beta}):
\bear
\alpha_V &=&  \, 2 \epsilon_V \, - \,  2(n-3) \, e_m^V
\, -  \,  4  \, \delta F_{(2)} \,
\Frac{\overline{M}_V^{\,\,2}}{
\overline{M}_S^{\,\,2}} \, -\, 6\, ,
\nn\\
\beta_V &=&  \, 2 \epsilon_V \, - \,  2(n-5) \,  e_m^V
\, -  \,  4 \,  \delta F_{(2)} \,
\Frac{\overline{M}_V^{\,\,2}}{
\overline{M}_S^{\,\,2}} \, -\, 6\, ,
\nn\\
\alpha_S&=& \, 2 \epsilon_S\, - \,4  \,  \delta F_{(2)}  \, - \, 2\, ,
\nn\\
\beta_S&=& \, 2 \epsilon_S\,+ \, 4 \,  e^S_m \, - \,4   \, \delta F_{(2)} \, - \, 2
\,.
\eear

Substituting the same experimental inputs as before
%%%in Eqs.~(\ref{splitV1})--(\ref{splitS}),
one obtains
\bear
\alpha_V = -7.5 \pm 0.3 \, ,
\qquad
\beta_V = -8.4 \pm 0.3\, ,
\qquad \mbox{for $n=5$,}
\nn\\
\alpha_V = -6.8 \pm 0.3\, ,
\qquad
\beta_V = -7.7 \pm 0.3\, ,
\qquad \mbox{for $n=3$.}
\eear
and for the scalar
\be
\alpha_S= 2\pm 7 \,, \qquad
 \beta_S = 2 \pm 8 \, .
\ee
Although $\alpha_R$ is not needed in the present $\cO(p^6)$ LEC study,
it is provided here for sake of completeness.

\subsection{NNLO chiral correction $\gamma_R$}

The parameter $\gamma_R$ -appearing
in $\Gamma_R/M_R^3$ at $\cO(m_\pi^4)$-  is even more complicate to determine
than $\alpha_R$ and $\beta_R$.

At next-to-next-to-leading order (NNLO), the resonance masses
may suffer  corrections due to the chiral operators
\be
\Frac{\we_{m,1}^R}{2 \overline{M}_R^{\,\,2}}\,  \bra RR \chi_+ \chi_+\ket
\,\,\, +\,\,\,
\Frac{\we_{m,2}^R}{2 \overline{M}_R^{\,\,2}} \, \bra R\chi_+ R \chi_+\ket\, ,
\ee
which combined with the leading and NLO operators in Eq.~(\ref{eq.eRm})
yield the pattern
\bear
M_{I=1}^2 &=& \overline{M}_R^2\,\, -\,\, 4 e^R_m m_{\pi}^2
\,\, -\,\, 4 (\we^{R}_{m,1}+\we^{R}_{m,2})
\Frac{m_{\pi}^4}{\overline{M}_R^{\,\, 2}}\,
%%%\,\,\, +\,\,\,\,\cO(m_{\pi}^6)
\, ,
\\
M_{I=1/2}^2 &=& \overline{M}_R^2\,\, -\,\, 4 e^R_m m_{ K}^2
\,\, -\,\,  \Frac{4 }{\overline{M}_R^{\,\, 2}}
\, \left[ \we^{R}_{m,1} \, (2 m_K^4 -2 m_K^2 m_\pi^2 + m_\pi^4)\,
+\, \we^{R}_{m,2} \, m_\pi^4\right]
%%%\, \,\,\,+\,\,\,\,\cO(m_{\pi}^6)
\, ,
\nn
\\
M_{I=0}^{(\bar{s}s)\,\,\,2} &=& \overline{M}_R^2\,\,
-\,\, 4 e^R_m (2 m_{ K}^2-m_\pi^2)
\,\, -\,\, 4 (\we^{R}_{m,1} + \we^{R}_{m,2}) \,
\Frac{(2 m_K^2-m_\pi^2)^2}{\overline{M}_R^{\,\, 2}}
\,
\, ,
\nn
\eear
and $M_{I=0}^{(\bar{u}u +\bar{d}d)}=M_{I=1}$.
Actually, the NNLO multiplet splitting will be only relevant for the analysis
of the vector parameter $\gamma_V$ in the Proca formalism ($n=5$).
In the remaining cases, $\we_{m}^R\equiv \we_{m,1}^R+\we_{m,2}^R$
will not appear. Since these NNLO quark mass corrections may enter in serious competition
with those NLO in $1/N_C$, an analysis of the vector spectrum in order to fix
$\widetilde{e}^V_m$  seems unreliable. Thus, we consider just some conservative bounds.
It is not really possible to obtain a tight constraint from
the $I=1$ or $I=\frac{1}{2}$.  The most stringent bound comes from
the $I=0$ $(\bar{s}s)$ state:
\be
\, \left| \we^{V}_{m} \right|
\,\, \leq \,\,
 \Frac{ \overline{M}_V^{\,\, 2}}{2 m_K^2-m_\pi^2}\,\,
 |e^V_m|  \,\,\,\,
 \simeq \,\,\,\, 0.3
\, ,
\ee
where we have demanded that the NNLO could not overcome the NLO  contribution
to the $\phi(1020)$ mass (in the quark mass expansion).

Likewise, the expansion of the $\rho$ and $\sigma$ resonance widths
up to NNLO in $m_\pi^2$ is given by the parameters $\widetilde{\epsilon}_R$:
\bear
\Gamma_{\rho\to\pi\pi} &=&
\Frac{ M_\rho^{\,\, n} \, \rho_{\rho\pi\pi}^3 }{48\, \pi \,  F_{\pi}^4}
\,\,\, \lambda_{V\pi\pi}^2\,\,\, \left[1 + \epsilon_V \Frac{m_\pi^2}{\overline{M}_V^{\,\, 2}}
\,
+ \,  \widetilde{\epsilon}_V\Frac{m_\pi^4}{\overline{M}_V^{\,\, 4}} \,\,\,+\,\,\, \cO(m_\pi^6)\right]^2\,,
\nn\\
\Gamma_{\sigma\to \pi\pi} \,\,&=&\,\,
\Frac{3 M_\sigma^{3} \rho_{\sigma\pi\pi}}{16\pi F_\pi^4} \,\,\,
c_d^2 \, \left[1  +  \epsilon_S \Frac{m_\pi^2}{\overline{M}_S^{\,\,2}}
+  \widetilde{\epsilon}_S\Frac{m_\pi^4}{\overline{M}_S^{\,\, 4}} + \cO(m_\pi^6)\right]^2\, .
\eear
A way out to estimate these chiral corrections is
the phenomenological lagrangian of  Eq.(\ref{lagr}):
\bqa
%%\widetilde{\epsilon}_V&=&\Frac{32\sqrt2\,f_\chi  c_m(c_d-c_m)
%%\overline{M}^2_V }{g_V F^2 \overline{M}^2_S}+
%%\Frac{16\sqrt2\,f_\chi e_m^V }{g_V }\,,
%%
\widetilde{\epsilon}_V&=&\epsilon_V \,\left[ \Frac{8  c_m(c_d-c_m)}{F^2}
\,\Frac{\overline{M}^2_V }{\overline{M}^{\,\, 2}_S}\, +
4 e_m^V \right]\,,
\nonumber \\
\widetilde{\epsilon}_S&=&\Frac{16 c_m^2(c_d-c_m)}{c_d F^2}
+ \Frac{8 (c_m-c_d) e_m^S }{c_d }\,,
\eqa
with $\epsilon_V= 4\sqrt{2} f_\chi/\lambda_{V\pi\pi}$
and the mass splitting operators~(\ref{eq.eRm}) also taken into account.
The full unexpanded expression for the widths can be found in
Ref.\cite{matching-op6}. Using the inputs of former sections, one obtains
\bear
\widetilde{\epsilon}_V&=&-0.0 \pm 0.3 \, \,  \mbox{(for n=5),}
\qquad\qquad\qquad
\widetilde{\epsilon}_V= -2.8 \pm 1.7  \,  \, \mbox{(for n=3),}
\nn\\
\widetilde{\epsilon}_S &=& -30 \pm 40\, .
\eear

Gathering all the different contributions together,
it is now possible to get  the NNLO correction $\gamma_R$ to
the ratio $\Gamma_R/M_R^3$:
\bqa
\gamma_V &=&
2 \widetilde{\epsilon}_V\, - 2(n-3) \we^V_m
- 4 \delta F_{(4)}
\Frac{\overline{M}_V^{\,\,4}}{\overline{M}_S^{\,\,4}}
\label{eq.gV}
\\
&&
+\epsilon_V^2 - 4 (n-3) e^V_m \epsilon_V - 8 \epsilon_V \delta F_{(2)}
\Frac{\overline{M}_V^{\,\,2}}{\overline{M}_S^{\,\,2}}
-12 \epsilon_V + 2 (n^2-8n +15) {e^V_m}^2
\nn \\
&&
+ 10 (\delta F_{(2)})^2
\Frac{\overline{M}_V^{\,\,4}}{\overline{M}_S^{\,\,4}}
+ 12 (n-5) e^V_m
+ 24 \delta F_{(2)}
\Frac{\overline{M}_V^{\,\,2}}{\overline{M}_S^{\,\,2}}
+ 8 (n-3) e^V_m \delta F_{(2)}
\Frac{\overline{M}_V^{\,\,2}}{\overline{M}_S^{\,\,2}}
+6\, ,
\nn\\
\nn\\
\gamma_S&=&
2 \widetilde{\epsilon}_S - 4 \delta F_{(4)}
\nn\\
&&
+\epsilon_S^2 - 4 \epsilon_S - 8 \epsilon_S \delta F_{(2)}
+ 8 \delta F_{(2)}+10 (\delta F_{(2)})^2
- 8 e^S_m -2\, .
\label{eq.gS}
\eqa
Notice that the $\cO(m_\pi^4)$ corrections to $F_\pi$  ($\delta F_{(4)}$),
the resonance masses ($\we_m^R$) and widths ($\widetilde{\epsilon}_R$)
are presented in the first line of
Eqs.~(\ref{eq.gV})--(\ref{eq.gS}).
The results of Eqs.~(\ref{eq.gV})--(\ref{eq.gS}) are actually
model dependent since there exist several other resonance operators
that may contribute to the
NNLO parameters $\delta F_{(4)}$, $\widetilde{\epsilon}_V$,
$\widetilde{\epsilon}_S$~\cite{op6rxt}.
%%%%, $\we_m^R$
Nevertheless, if one relies in the model of Eq.~(\ref{lagr}) the inputs
considered in former sections lead to the chiral symmetry breaking terms
\bqa
\gamma_V &=& 30 \pm 10\quad  \mbox{(for n=5),}\qquad\qquad \qquad
\gamma_V = 19 \pm 8\quad \mbox{(for n=3)},
\nn\\
\gamma_S &=& -50\pm 70\, .
\eqa
A more detailed observation of the  NNLO parameters
(first line of Eqs.~(\ref{eq.gV})--(\ref{eq.gS}))
shows that the impact of the
NNLO mass splitting term $\widetilde{e}^R_m$ is negligible.
On the other hand, $\delta F_{(4)}$ and $\widetilde{\epsilon}_S$
are responsible of most
of the error in $\gamma_S$. In the vector case, the effect of the width splitting term
$\widetilde{\epsilon}_V$ is not dominant, whereas the $F_\pi$
NNLO correction $\delta F_{(4)}$ is responsible of roughly the 75\%
of the total uncertainty in $\gamma_V$.  The large errors
we have in our input  numbers for $\delta F_{(4)}$ and $\widetilde{\epsilon}_S$
reassure us in the validity of our result, even if the uncertainty of the $\gamma_R$
covers a rather conservative interval.

\section{Low-energy constant determination at $\cO(p^6)$}

\subsection{The determination of $r_5$, $r_6$}
\label{sec.r5r6}

The couplings  $r_5, r_6$  only depend
on the chiral limit values of the resonance masses and decay widths~\cite{matching-op6}
and, therefore, are the most reliably determined low-energy constants:
\be
\label{r5}
r_5=\frac{32 \pi F^6 \overline{\Gamma}_\sigma}{3\overline{M}_S^{\,\,7}}
+\frac{36 \pi F^6 \overline{\Gamma}_\rho}{\overline{M}_V^{\,\,7}}
\, ,
\ee
\be \label{r6}
r_6=\frac{12 \pi F^6 \overline{\Gamma}_\rho}{\overline{M}_V^{\,\,7}} \,.
\ee

\subsection{Extraction of $r_3$, $r_4$}

In the case of the couplings $r_3$ and $r_4$, all one needs are
the resonance widths and masses in the chiral limit and
the first $m_\pi^2$ correction to the ratio $\Gamma_{\rm R}/M_{\rm R}^5$,
this is,  $\beta_{\rm R}$~\cite{matching-op6}:
\be \label{r3}
r_3=\frac{64 \pi F^6 \overline{\Gamma}_\sigma}{3\overline{M}_S^{\,\,7}} \left(
1+\frac{\beta_{\rm S}}{2}\right) \, - \, \frac{768 \pi F^6
\overline{\Gamma}_\rho}{\overline{M}_V^{\,\,7}} ( 1 + \frac{3  \beta_{
V}}{32} )  \, ,
\ee

\be
\label{r4}
 r_4=\frac{192 \pi F^6 \overline{\Gamma}_\rho}{\overline{M}_V^{\,\,7}}
 \left( 1 +\frac{\beta_{V}}{8}\right)
\, .
\ee
The coupling $r_4$ depends only on the vector resonance.
Likewise, the scalar contribution is quite suppressed in $r_3$
due to the large numerical coefficient in front of the vector term.

\subsection{Extraction of $r_2$}
\label{sec.r2}

Compared to the determination of $r_3,\,...r_6$, the coupling
$r_2$ carries larger theoretical uncertainties
since the NNLO parameters $\gamma_S$ and $\gamma_V$ enter into play.
The prediction for $r_2$ derived from partial-wave dispersion
relations is given by~\cite{matching-op6}
\bqa \label{r2rf}
 r_2 & = & 2 r_F\,\, +\,\,
 \frac{64 \pi F^6 \overline{\Gamma}_\sigma}{\overline{M}_S^{\,\,7}} \left( 1+
\frac{\beta_{\rm S}}{3} +\frac{\gamma_{\rm S}}{6}   \right) +\frac{\pi F^6
\overline{\Gamma}_\rho}{\overline{M}_V^{\,\,7}} \left( 7584 + 1248 \beta_{\rm
V} + 144 \gamma_{\rm V} \right) \,, \nonumber \\
\eqa
where  $r_F$ provides $F_\pi$ at NNLO in $m_\pi^2$~\cite{bijff}.
It is related to our previously defined $\delta F_{(4)}$ through
$r_F=2 \ell_3 \ell_4 +  F^4\delta F_{(4)}/\overline{M}_S^{\,\, 4}$.
The phenomenological lagrangian~(\ref{lagr}) produces  the value of $\delta F_{(4)}$
in Eq.~(\ref{eq.Fpi-ecker}), which in combination with
${    2\ell_3\ell_4= 32 c_m^2 c_d (c_m-c_d)/\overline{M}_S^{\,\,4}     }$~\cite{ecker89} yields
\bqa
r_F=-\frac{8 c_d^2 c_m^2}{\overline{M}_S^{\,\,4}}
+\frac{16 e_m^S c_d c_m F^2}{\overline{M}_S^{\,\,4}}\,.
\eqa
Substituting our former phenomenological inputs, one gets
$r_F=(-1\pm 3) \cdot 10^{-4}$.

\subsection{Extraction of $r_1$ and $r_{S2}$}
Within the framework of $\pi\pi$ partial-wave sum-rules proposed in
Refs.~\cite{matching,matching-op6}, it is not possible to make any prediction for
$r_1$, as they are  based on  once-subtracted dispersion relation
and this coupling produces just a constant contribution to the $\pi\pi$--scattering
amplitude.

The couplings $r_1$ (from   $\pi\pi$--scattering) and
$r_{S2}$ (from the $\pi\pi$ scalar form-factor) must be extracted through alternative
procedures.    The values from Eqs.~(\ref{r1nosplitana})
and (\ref{rs2nosplitana}) were based on the  lagrangian~(\ref{lagr}) but
without accounting for the mass splitting effect.
If the latter is included,  our phenomenological model produces
for $r_1$ the new predictions
\bear
r_1^{\rm n=5}&=&  -\Frac{16 c_d c_m ( 8 c_d^2-17 c_d c_m + 12 c_m^2)}{\overline{M}_S^{\,\,4}}
+ \Frac{32(c_d-c_m)^2 F^2}{\overline{M}_S^{\,\,4}}\, e_m^S
\nn\\
&&\qquad
-\Frac{16 g_V^2 F^2}{\overline{M}_V^{\,\,2} }
\left[1 + \epsilon_V +\Frac{1}{4}  \epsilon_V ^2
- \Frac{ 8 c_d c_m}{F^2} \Frac{\overline{M}_V^{\,\,2}}{\overline{M}_S^{\,\,2}} \right]
\, ,
\nn\\
r_1^{\rm n=3}&=&  -\Frac{16 c_d c_m ( 8 c_d^2-17 c_d c_m + 12 c_m^2)}{\overline{M}_S^{\,\,4}}
+ \Frac{32(c_d-c_m)^2 F^2}{\overline{M}_S^{\,\,4}}\, e_m^S
\nn\\
&&\qquad
-\Frac{16 G_V^2F^2}{\overline{M}_V^{\,\,4}} \left[ 1+ \epsilon_V
- \Frac{ 8 c_d c_m}{F^2} \Frac{\overline{M}_V^{\,\,2}}{\overline{M}_S^{\,\,2}}
+ 2 e^V_m  \right] \, .
\nn\\
\label{eq.r1+split}
\eear
%%%
%%%\bqa
%%%n=5 \longrightarrow\quad r_1^{n=5}  &=&-\frac{4 g_V^2F^2}{M_V^2} \left(2+ \frac{4\sqrt{2}f_\chi}{g_V}\right)^2
%%%\nonumber \\
%%%&& -\Frac{16 c_d c_m ( 8 c_d^2-17 c_d c_m + 12 c_m^2)}{M_S^4}+ \Frac{32(c_d-c_m)^2 F^2}{M_S^4}e_m^S
%%%\,,
%%%\\
%%%\nn\\
%%%n=3 \longrightarrow\quad r_1^{n=3} &=&r_1^{n=5}\,\,\,
%%%+\,\,\, \Frac{128 f_\chi^2 F^2}{\overline{M}_V^{\,\,4}}\,.
%%%\eqa
The detailed calculation is relegated to Appendix~\ref{app.r1}.

Likewise, after taking into account the mass splitting in the scalar form-factor
calculation (Appendix~\ref{app.SFF}),  $r_{S_2}$ becomes
\bqa
r_{S2} &=&  \Frac{ 8 c_m (c_m-c_d) F^2}{\overline{M}_S^{\,\,4}}
\,-\, \Frac{32 c_d^2 c_m^2}{\overline{M}_S^{\,\,4}}
+ \frac{16c_d c_m F^2}{\overline{M}_S^{\,\,4}}e_m^S\,.
\label{eq.rs2+split}
\eqa

\subsection{Saturation scale uncertainty}

One last problem to face is the fact that the large--$N_C$ estimate of the  LECs
does not carry any renormalization scale dependence.
However, it is possible to find a so called ``saturation scale'' $\mu_s$
such that the physical LEC $r_i^r(\mu)$ agrees numerically
with $r_i^{N_C\to\infty}$ for $\mu=\mu_s$.

Since the standard comparison  scale is $\mu_0=770$~MeV,
the coupling $r_i^r(\mu)$ must be run from $\mu_s$ up to $\mu_0$.
The possible  difference between these two scales  introduces an uncertainty
in our determination. The way considered here to account for this lack of knowledge
is to observe the variation for a wide range of $\mu$, which
in other works is usually taken to be in the
range 500--1000~MeV~\cite{slgasser,bijri,op6rxt,SPP}:
\bqa\label{rimu}
&&{\Delta r_1}^{\mu_s} = 3 \times 10^{-4}\,,
\quad {\Delta r_2}^{\mu_s} = 4\times 10^{-4}\,,\quad
{\Delta r_3}^{\mu_s} = 3\times 10^{-4}\,, \nonumber\\ &&
{\Delta r_4}^{\mu_s} = 0.05\times 10^{-4}\,,\quad
{\Delta r_5}^{\mu_s} = 0.5\times 10^{-4}\,,
\quad {\Delta r_6}^{\mu_s} = 0.05\times 10^{-4}\,,
\nonumber \\&&
{\Delta r_{S2}}^{\mu_s} = 1.5\times 10^{-4}\,,
\eqa
where the running is completely fixed in $\chi$PT by the expressions given
in Appendix~B in terms of $F$ and the $\cO(p^4)$ invariants
$\bar{\ell}_1=-0.4\pm 0.6$, $\bar{\ell}_2=4.3\pm 0.1$,
$\bar{\ell}_4=4.4\pm 0.2$~\cite{slgasser} and
$\bar{\ell}_3=2.9\pm 2.4$~\cite{gasser84}.

Sometimes, it is argued that there must exist a common saturation scale
for all the LECs, both $\cO(p^4)$ and $\cO(p^6)$. This scale $\mu_s$  is
very often identified with the rho mass, $\mu_s=M_\rho$.
However, explicit one-loop calculations with resonance lagrangians
show that, though the loops typically produce logarithms of the form
$\ln\frac{M_R}{\mu}$ or $\ln\frac{M_R}{M_R'}$, the combinations of
them appearing for each LEC
are not necessarily the same~\cite{Cata+Peris,L8,L9,L10}.
Hence,  although in general terms $\mu_s\sim M_R$,
nothing ensures that the saturation scales must be exactly identical,
nor that their value  must be equal to $M_\rho$~\cite{L10,Kampf-L2}.
Thus, the uncertainty of every LEC is computed and accounted separately
in the present work.

\subsection{Summary of low-energy constants}

The values for the low-energy constants $r_i^r(\mu)$ and the corresponding errors are
gathered  in Table~\ref{tab.ri}.  We choose the standard comparison scale $\mu=770$~MeV
and include the uncertainty from the saturation scale estimated in the former section
as an error.

\begin{table}[t]
\centering
\begin{tabular}{|c|c|c|c|c|}
  \hline
 & ND est. & set A  & set C  & set C
\\
 &  \cite{op6-reno} & \cite{bijri,bijff}  &   (n=5) &  (n=3)
   \\
    \hline

  $10^4 \cdot r_1^r$ & $\pm 80$ &$-0.6$  &
  %%%$-24 \pm 8 \pm 3 $
  $-14\pm 17\pm 3$  &
  %%%$-36 \pm 8 \pm 3$
  $-20\pm 17 \pm 3$
  \\
  $ 10^4 \cdot r_2^r$& $\pm 40$ & $1.3$ &
  $22\pm 16\pm 4  $   &  $ 7\pm 10\pm 4 $
  \\
  $10^4 \cdot r_3^r$ & $\pm 20$ & $-1.7$ &
  $-3\pm 1 \pm 3$    & $ -4\pm 1  \pm 3$
  \\
  $10^4 \cdot r_4^r$ & $\pm 3$ &$-1.0$ &
  $-0.22\pm 0.13\pm 0.05$     & $ 0.13\pm 0.13\pm 0.05$
  \\
  $10^4 \cdot r_5^r$ & $\pm 6$ & $1.1$  &
  $0.9\pm 0.1\pm 0.5 $     & $0.9\pm 0.1 \pm 0.5$
  \\
  $10^4 \cdot r_6^r$ & $\pm 2$ & $0.3$  &
  $0.25\pm 0.01\pm 0.05$    & $ 0.25\pm 0.01\pm 0.05$
  \\
  $10^4 \cdot r_{S_2}^r$ & $\pm 1$ & $-0.3 $  &
  $1\pm 4 \pm 1 $    & $1\pm 4 \pm 1 $
  \\
  \hline
\end{tabular}
\caption{{\small
Different predictions for the $\cO(p^6)$ LECs  $r_i^r(\mu)$ for $\mu=770$~MeV:
The first column presents the order of magnitude estimate based on
naive dimensional analysis~\cite{op6-reno};
In the set A column  we show former estimates from Refs.~\cite{bijri,bijff};
in the last two columns,
one can find the values for the present reanalysis, both in
the Proca (n=5) and antisymmetric vector formalism (n=3).
The first error derives from the inputs and the second from the uncertainty
in the saturation scale.}}
\label{tab.ri}
\end{table}

Some remarks about the present predictions are in order.
The LECs $r_5$ and $r_6$ are the most model-independent ones
since they are determined  by just the chiral limit of the resonances
masses and widths (see Eqs.(\ref{r5}) and (\ref{r6})).
Our determinations for $r_5$ and $r_6$ are consistent
with those in Ref.~\cite{bijri}.  The constants $r_3$ and $r_4$ depend on  the
$\cO(m_\pi^2)$ corrections $\beta_R$ in the ratio $\Gamma_R/M_R^5$ ,
which are still rather under control.
All our predictions are in agreement within errors with  those
reported in Ref.~\cite{bijri} except for the small deviation found for  $r_4$,
mainly due to the slight discrepancies in the coupling $f_\chi$
mentioned in previous sections.   All this points out the little model dependence
of our LEC estimate.

On the other hand, $r_2$ is partly determined by the
NNLO $m_\pi^2$ corrections $\gamma_R$ to the ratio $\Gamma_R/M_R^3$.
This makes it one of the least controlled LECs.
Roughly half of the total error of the LEC is due to the NNLO $m_\pi^2$ corrections
$\delta F_{(4)}$ and $\widetilde{\epsilon}_S$ in $\gamma_S$.
Thus, the uncertainty of $r_2$ is found to be dominated by the scalar mass
parameters, which is due in our case to our poor knowledge of the
$I=1/2$ scalar mass.

In general, the value of all the couplings that can be  determined by means of
the partial wave dispersion relations  ($r_2$,... $r_6$)~\cite{matching,matching-op6}
is found to be dominated by the vector contributions.
Since this sector is rather under control, this reassures us in the
robustness of our calculation.

Finally, the couplings $r_1$ and $r_{S_2}$  need to be estimated through a
phenomenological model. In the case of the lagrangian~(\ref{lagr})
the error of $r_1$ is mainly due to our ignorance  on the scalar resonances
couplings $c_d$ and $c_m$. This is directly originated in the large
uncertainty of the $a_0\to \pi\eta$ decay width.
The vector contribution is much smaller and essentially negligible.  The error in $r_{S_2}$ stems
mainly from our poor knowledge of the $a_0(980)$ width and the scalar mass splitting
(which derives here from the large uncertainty of the $I=1/2$ scalar resonance).

In summary, we find in general a good agreement with former determinations and at the same time we are able to provide a reliable estimate of the error.
This will help us to establish  the relevance of the $\cO(p^6)$ LEC
contributions to the scattering lengths in next section.

\section{Scattering lengths }

\subsection{Direct $\chi$PT calculation}

\begin{table}[ht]
\centering
\begin{tabular}{|c|c|c|c|}
  \hline
 & Set A~\cite{bijri}  & Set C (n=5) &  Set C (n=3)
   \\
  & $\qquad ( \times 10^{-3} ) \qquad $  &  $ ( \times 10^{-3} )$  &
  $ ( \times 10^{-3} )$
   \\
    \hline
     \hline
  $a^0_0$ & $1.2$  &  $2.3 \pm 1.3 \pm 1.7 $  & $-0.3 \pm 1.0 \pm 1.7$
  \\
  $ b^0_0$&  $4.1$ & $4 \pm 1 \pm 4 $   &  $  2 \pm 1 \pm 4$
  \\
  $10 a^2_0$ & $-3.7$ &   $-4 \pm 4 \pm 1$    & $ -4\pm 4 \pm 1$
  \\
  $10  b^2_0$ & $14$ & $ -13 \pm 8 \pm 2 $     & $  -3 \pm 5 \pm 2$
  \\
  $10 a^1_1$ & $-0.3$  &  $ 3.4 \pm 2.2 \pm 0.6 $     & $ 1.7 \pm 1.3  \pm 0.6$
  \\
  $10  b^1_1$ & $0.7$  &   $2.6 \pm 0.5 \pm 1.5$    & $ 3.7 \pm 0.6 \pm 1.5$
  \\
  \hline
\end{tabular}
\caption{{\small
Contribution of the $\cO(p^6)$ LECs $r_i^r(\mu)$
to the scattering lengths $a^I_J$ and effective
ranges $b^I_J$ in the direct $\chi$PT approach~\cite{bijri}
for $\mu=770$~MeV.
The first column (set A) shows the results
from Ref.~\cite{bijri}, with their same inputs
$F_\pi=93.2$~MeV, $m_\pi=139.47$~MeV.
The last two columns  show the predictions based on our phenomenological
reanalysis  (set C) for the Proca (n=5) and antisymmetric tensor formalisms (n=3).
The first error derives from the inputs and the second one from the
saturation scale uncertainty.
}} \label{tab.daIJ-Bij}
\end{table}

The $\pi\pi$--scattering lengths were first calculated in $\chi$PT up to $\cO(p^6)$
in Ref.~\cite{bijri}.  At order $m_\pi^6$,
the contribution  from the $r_i^r(\mu)$ LECs was found to be
\bqa
a_0^0|_{r_i}&=& \frac{m_\pi^6}{32 \pi F_\pi^6}
\left[5 r_1^r + 12  r_2^r +48  r_3^r +32  r_4^r + 192 r_5^r   \right]\,,\nonumber\\
b_0^0|_{r_i}&=& \frac{m_\pi^6}{4 \pi F_\pi^6}
\left[ r_2^r +12  r_3^r +12  r_4^r + 72 r_5^r - 8 r_6^r \right]\,,\nonumber\\
a_0^2|_{r_i}&=& \frac{m_\pi^6}{16 \pi F_\pi^6}
\left[  r_1^r + 16 r_4^r \right]\,,\nonumber\\
b_0^2|_{r_i}&=& \frac{m_\pi^6}{8 \pi F_\pi^6}
\left[- r_2^r +24 r_4^r - 16 r_6^r\right]\,,\nonumber\\
a_1^1|_{r_i}&=& \frac{m_\pi^6}{24 \pi F_\pi^6}
\left[  r_2^r  +8 r_4^r + 16 r_6^r \right]\,,\nonumber\\
b_1^1|_{r_i}&=& \frac{m_\pi^6}{6\pi F_\pi^6}
\left[ -  r_3^r +3  r_4^r + 8 r_6^r \right]\,.
\label{eq.aIJ-Bij}
\eqa
In order to work with dimensionless  quantities,
we have multiplied the results in Ref.~\cite{bijri}
for $b_0^0$, $b_0^2$, $a_1^1$ by  $m_\pi^2$, and for $b_1^1$ by $m_\pi^4$.
Our estimate of the contributions to the scattering lengths and effective ranges
for the standard comparison scale ${   \mu=770    }$~MeV is
given in Table~\ref{tab.daIJ-Bij}, both for the Proca (n=5)
and the antisymmetric vector formalism (n=3).
In order to have a clearer comparison of this quantities, the outcome from Ref.~\cite{bijri}
is also provided.  The first error in Table~\ref{tab.daIJ-Bij}
comes from the phenomenological inputs and the second one from the saturation scale
uncertainty in the $\cO{(p^6)}$ LECs, given in Eq.(\ref{rimu}).

An important part of the subdominant quark mass corrections
has been pinned down quite accurately through the comparison of the decays
of the different resonances in the multiplet.
However, there exist
a series of new operators (e.g. the vector resonance operator $\lambda^V_9$
in Ref.~\cite{op6rxt}) whose couplings  cannot be extracted in an independent
way. If present, they could enter in effective combinations $g_V(m_\pi)^{\rm eff}$
and $c_d(m_\pi)^{\rm eff}$; these would be what we would really determine and denote
respectively as $g_V$ and $c_d$, being used later instead of them in the LEC
computation.
A more detailed discussion is relegated to Appendix~\ref{lambdaV8}.
In any case, the study of these operators remains beyond the scope of this article.

\subsection{CGL dispersive method}

In Ref.~\cite{slgasser}, Colangelo {\it  et al.} combined the NNLO chiral
perturbation theory computation of the scattering lengths~\cite{bijri}
with a  phenomenological dispersive representation.  This allowed them to produce
one of the most precise determinations of the scattering lengths.
These were expressed in terms of some dispersive integrals,
the pion quadratic scalar radius $\langle r^2\rangle_S^\pi$,
the $\cO(p^4)$ coupling $\ell_3$ and a set of $\cO(p^6)$ LECs
($r_1,...\, r_4$, $r_{S_2}$).   These last contributions are actually
the most poorly known and the aim of this paper.
Following the work of Ref.\cite{slgasser},
we extract the part of their scattering lengths  that depends on the inputs  $r_i^r(\mu)$:
\bqa
a^0_0|_{r_i} =& \frac{7m_\pi^2}{32\pi F_\pi^2}  C_0|_{r_i}\, &=\,
\Frac{m_\pi^6}{32 \pi F_\pi^6}
\left[ 5 r_1^r + 12 r_2^r + 28 r_3^r - 28 r_4^r -14 r_{S_2}\right] \, ,\quad
\nonumber \\
a^2_0|_{r_i}  =& -\frac{m_\pi^2}{16\pi F_\pi^2}  C_2|_{r_i}
\,&=\, \Frac{m_\pi^6}{16\pi F_\pi^6}
\left[ r_1^r - 4r_3 + 4 r_4 + 2 r_{S_2}\right]
\,,
\label{scalgasser}
\eqa
where the $C_j|_{r_i}$  can be extracted from the $C_j$ provided
in the Appendix~C of Ref.~\cite{slgasser}.
%%%:
%%%\bqa
%%%C_1|_{r_i} &=& \frac{m_\pi^4}{F_\pi^4}( r_2^r + 4 r_3^r -4  r_4^r -2 r_{S2}^r )\,,
%%%\nonumber \\
%%%C_2|_{r_i} &=& \frac{m_\pi^4}{F_\pi^4}(-r_1^r + 4  r_3^r -4 r_4^r -2  r_{S2}^r) \,,
%%%\nonumber \\
%%%C_0|_{r_i} &=& \Frac{12}{7} C_1|_{r_i} -\Frac{5}{7} C_2|_{r_i} \,.
%%%\eqa
%where $\Delta r_i = r_i^{our}-r_i^{Bij}$ and $r_i^{our}$ are the values determined from the scenario C, i.e. the values
%presented in Eq.(\ref{r56sc}), Eq.(\ref{r34sc}), Eq.(\ref{r2sc}), Eq.(\ref{r1sc}) and Eq.(\ref{rs2sc}).
The LEC $r_{S_2}$ appears in this analysis~\cite{slgasser} because
the scalar radius $\bra r^2\ket_S^\pi$ is incorporated as an experimental information
in order to fix the $\cO(p^4)$ constant $\ell_4$.
On the other hand, the couplings $r_5$ and $r_6$ disappear here
with respect to the standard $\chi$PT analysis
of Ref.~\cite{bijri}, being their information encoded
and replaced by the different  dispersive integrals.
It is easy to realize that the resonance contributions
$a^I_J|_{r_i}$ carry the same $r_1$ and $r_2$ dependence in both the
direct $\chi$PT calculation in Eq.~(\ref{eq.aIJ-Bij})~\cite{bijri}
and the dispersive study in Eq.~(\ref{scalgasser})~\cite{slgasser}.
The comparison  of this reanalysis and  the predictions from Ref.~\cite{slgasser},
is shown in Table~\ref{tab.CGL} for the standard reference scale $\mu=770$~MeV.

\begin{table}[ht]
\centering
\begin{tabular}{|c|c|c|c|c|}
  \hline
 & Total: Ref.~\cite{slgasser}  &  $a^I_J|_{r_i}$~\cite{slgasser}
 & Set C \ \  (n=5) & Set C  \ \  (n=3)
   \\
  & $\qquad ( \times 10^{-3}) \qquad $ & $\qquad ( \times 10^{-3}) \qquad $
  &  $ ( \times 10^{-3})$  & $(\times 10^{-3})$
   \\
    \hline
     \hline
  $a^0_0$ &  $220\pm 5$ & $0.0\pm 1.0 $  &  $1.0 \pm 1.5 \pm 1.0 $
  & $-1.6 \pm 1.5 \pm 1.0$
  \\
  $10 a^2_0$ & $-444\pm 10$  & $0.4\pm 2.0 $ &   $0 \pm 4\pm 2$
  & $0\pm 4\pm 2$
  \\
  \hline
\end{tabular}
\caption{{\small   The first and second columns  show, respectively,
the total scattering lengths
and the $r_i$ contribution to them in the dispersive method from
Colangelo {\it et al.}~\cite{slgasser},
where the authors used the $r_i$ in Eq.~(2.3),
 $F_\pi=92.4$~MeV and $m_\pi=139.57$~MeV.
The last two columns  show the reanalyzed quantities (set C)  for
the Proca (n=5) and antisymmetric formalisms (n=3)
for the usual scale $\mu=770$~MeV. There,
the first error derives from the inputs and the second one from the
saturation scale uncertainty.
}}
\label{tab.CGL}
\end{table}

We find that the largest contributions to the $a_0^0$ and $a^2_0$ errors are produced in similar terms by
$r_1$, $r_2$, $r_3$ and $r_{S_2}$.  On the other hand, the impact of $r_4$
on both the value and uncertainty of the scattering lengths results negligible.

\section{Summary and conclusion}

This article concludes the previous work from Ref.~\cite{matching-op6}.
The analysis of the uncertainties of the $\cO(p^6)$ LECs
contributing to  the $\pi\pi$ scattering lengths~\cite{slgasser,bijri}
is completed here. Former estimates based on the large--$N_C$ limit
and resonance saturation  have been revised.
Nevertheless,  a series of uncertainties escape to the control of
the analysis carried in the present article.  Most of the computations
are rather model independent as they are based on general resonance properties;
however, at some points we had to rely on
phenomenological lagrangians~\cite{bijri,ecker89,mass-split}.

All this allowed the estimate of the $\cO(p^6)$ LECs.
For the standard comparison scale $\mu=770$~MeV, one obtains
after combining the results from Proca and antisymmetric formalism
the values
\bear
r_1^r=(-17\pm 20) \times 10^{-4}\, , \qquad  &
r_2^r=(17\pm 21)\, \times 10^{-4}\, , \quad
&r_3^r=(-4\pm 4)\,  \times 10^{-4}\, ,
\nn \\
r_4^r=(0.0\pm 0.3)\,  \times 10^{-4}\, , \qquad  &
r_5^r=(0.9 \pm 0.5 )\,  \times 10^{-4}\, , \quad
&r_6^r=(0.25 \pm 0.05)\, \times 10^{-4}\, ,
\nn\\
r_{S_2}^r=(1\pm 4)\,  \times 10^{-4}\, . \qquad  &
\eear

The combination of the results from  the Proca (n=5) and antisymmetric formalism
(n=3) yield for the dispersive method~\cite{slgasser}
the final resonance contribution,
\bqa
10^3 \, a^0_0|_{r_i}\, =\, 0 \pm 3 \, ,
\qquad\qquad
10^4  a^2_0|_{r_i} \, =\, 0 \pm 5 \, .
\label{eq.aIJ_ri}
\eqa
This is in  perfect agreement with the former estimate~\cite{slgasser},
although the detailed analysis of the phenomenological inputs
casts a slightly larger uncertainty.

Following the analysis of global uncertainties
of the scattering lengths in Colangelo {\it et al.}'s~\cite{slgasser},
we have verified that the global uncertainties for $a^0_0$ and $a^2_0$
are not largely modified.
Thus, the replacement of the values from Eq.~(\ref{eq.aIJ_ri})
in the total scattering lengths leads to the updated predictions
\bqa
a^0_0\, =\, 0.220 \pm 0.005\, ,
\qquad\qquad
10 a^2_0\, =\, -0.444\pm 0.011 \, , \,
\eqa
where the total uncertainties and central values
are essentially unchanged with respect
to the previous determinations  $a_0^0=0.220\pm 0.005$ and
$10 a^2_0=-0.444\pm 0.010$~\cite{slgasser}.

This work  provides a reliable and solid estimation
of this part of the $\cO(p^6)$ calculation.
It sets clear limits to the size of the $\cO(p^6)$ LEC contributions,
slightly conservative in some occasions.
A special attention has been put on the quantification of errors
and their precise source. It reassures us in the reliability
and precision of the current scattering length determination
based on the dispersive approach of Ref.~\cite{slgasser}.
Nonetheless, although the global uncertainties of the
scattering lengths are barely affected by our new predictions of
the $\cO(p^6)$ low energy constants, the total errors
have almost reached those from $a^I_J|_{r_i}$.
This points out the difficulty of further improvements in
the accuracy of the $\pi\pi$ scattering lengths
unless the $\cO(p^6)$ LEC uncertainties are conveniently reduced.

\section*{Acknowledgements}

This work has been supported in part by
CICYT-FEDER-FPA2008-01430, SGR2005-00916, the Spanish Consolider-Ingenio 2010
Program CPAN (CSD2007-00042), the Juan de la Cierva program,
the EU Contract No. MRTN-CT-2006-035482
(FLAVIAnet) and National Nature Science Foundation of China under grant number
   10875001, 10575002 and 10721063.
We are in debt with R.~Escribano for his comments on how to improve the treatment of the $\eta$ meson in the $a_0\to\eta\pi$ decay.

\appendix

\section{Full calculation based on the resonance lagrangian}

\subsection{Extraction of $r_1$: $\pi\pi$ scattering }
\label{app.r1}

The resonance lagrangian from Eq.~(\ref{lagr})~\cite{bijri,ecker89}
has been used to calculate the $\pi\pi$--scattering amplitude at
large $N_C$. The vector resonances were described there by means of Proca four-vector fields
$\hat{V}_\mu$. In addition, we also took into account the vector and scalar
mass splittings.  The amplitude
$\pi^+(p_1)\pi^-(p_2) \rightarrow \pi^0(p_3) \pi^0(p_4)$ results then
\bqa
\label{eq.scatProca}
A(s,t,u)^{\rm Proca}&=& \Frac{s-m_\pi^2}{ F_\pi^2}
\,+\,  \Frac{2}{F_\pi^4 (M_{\sigma}^2-s)} \left[ c_d s - 2 c_d m_{\pi}^2 + 2 c_m m^2 \right]^2
\\
&& \hspace*{-0.5cm}+ \Frac{u-s}{F_\pi^4 (t-M_\rho^2 )}
\left[g_V t + 4\sqrt{2} f_{\chi} m^2\right]^2
%%\nn\\
%%&& \quad
+ \Frac{t-s}{ F_\pi^4 (u-M_\rho^2)}
\left[ g_V u + 4\sqrt2 f_{\chi} m^2\right]^2\, .
\nn
\eqa
with $s=(p_1+p_2)^2$, $t=(p_1-p_3)^2$, $u=(p_1-p_4)^2$. We denote by $\rho$ and $\sigma$
respectively the $I=1$ vector and the $I=0$ scalar with $\bar{u}u+\bar{d}d$ content.
$F_\pi\simeq 92.4$~MeV is the pion decay constants, which deviates from
its chiral limit $F$ in the way prescribed in Eq.~(\ref{Fpi-F})
due to the scalar resonance tadpole in the $c_m$ operator of
the resonance chiral theory lagrangian~(\ref{lagr})~\cite{matching-op6,juanjotadp}.
The same happens with the physical pion mass $m_\pi^2$ and its value at leading order
in the chiral expansion, $m^2$, which are related through
\begin{eqnarray}
\label{tadpole}
%%%F_\pi &=& F\,\,  \left[1 +\Frac{4 c_d c_m}{F^2}\, \Frac{m_\pi^2}{M_S^2}
%%%+\Frac{8 c_d c_m^2( 3 c_d-4 c_m) }{F^4} \,\Frac{m_\pi^4}{M_S^4}
%%%+\cO\left(\frac{m_\pi^6}{M_S^6}\right)\,\right]\, ,
%%%\nn\\
m^2\equiv
2 B_0 m_{_{u/d}}  &= &  m_\pi^2 \, +\, \Frac{8 c_m (c_d-c_m)}{f^2}\,
\Frac{m_\pi^4}{\overline{M}_S^2} \, +\,
\cO(m_\pi^6)\, ,
\label{eq.Fpi+mpi}
\end{eqnarray}
with $m_{_{u/d}}$ the $u/d$ quark mass in the isospin limit.

Alternatively, if one employs the antisymmetric tensor formalism
to describe the vector resonances~\cite{ecker89,op6rxt,gasser84}
the scattering takes the form
\bqa
A(s,t,u)^{\rm Antis.}&=& \Frac{s-m_\pi^2}{ F_\pi^2}
\,+\,  \Frac{2}{F_\pi^4 (M_{\sigma}^2-s)} \left[ c_d s - 2 c_d m_{\pi}^2 + 2 c_m m^2 \right]^2
\nn\\
&& + \Frac{u-s}{F_\pi^4 (t-M_\rho^2 )}
t \, \left[G_V  + 4\sqrt{2} f_{\chi}^{\rm n=3} m^2/M_\rho^2\right]^2
%%\nn\\
%%&& \quad
+ \Frac{t-s}{ F_\pi^4 (u-M_\rho^2)}
u\, \left[ G_V + 4\sqrt2 f_{\chi}^{\rm n=3} m^2/M_\rho^2\right]^2\,,
\label{eq.scatAnti}
\eqa
where the meaning of $G_V$ and $f_\chi^{\rm n=3}$ will be discussed later in detail in Appendix~\ref{lambdaV8}.

The resonance expressions can be now compared to the $\chi$PT amplitude~\cite{bijri},
\begin{eqnarray}
\label{op6ampop2}
A(s,t,u)^{\rm \chi PT}\, = && \,  \frac{s-m_\pi^2}{F^2}
+\frac{m_\pi^4}{F^4}\left( 8\ell_1 + 2\ell_3 \right)
- \frac{ 8 m_\pi^2s}{F^4}\ell_1
+\frac{s^2}{F^4}(2\ell_1+\frac{\ell_2}{2}) \nonumber
\\ &&
\,
+\frac{(t-u)^2}{2F^4}\ell_2
-\frac{8 m_\pi^6}{F^6}\ell_3^2
+\frac{m_\pi^6}{F^6} \left(r_1+ 2 r_F\right)
+\frac{m_\pi^4s}{F^6} \left(r_2- 2 r_F\right)
\nonumber
\\ &&
\, + \,  \frac{m_\pi^2s^2}{F^6}r_3
+ \frac{m_\pi^2(t-u)^2}{F^6}r_4
+\frac{s^3}{F^6}r_5
+\frac{s(t-u)^2}{F^6}r_6 \,. \nonumber
\\ &&
\end{eqnarray}
We have preferred to express everything in terms of $F$ rather than $F_\pi$
in order to make  the chiral matching more transparent~\cite{matching,matching-op6}.

The LECs are extracted by matching  $A(s,t,u)^{\rm \chi PT}$ and the chiral expansion
of the resonance amplitude $A(s,t,u)^{\rm Res}$ in Eqs.~(\ref{eq.scatProca}) and
(\ref{eq.scatAnti}).
Since we are interested in $r_1$, $s=t=0$ and $ u=4m_\pi^2$ turn out to be the most convenient choice of
momenta. The matching of the $m_\pi^6$ term yields for the Proca formalism (n=5)
\bqa
r_1   +2r_F +16 r_4-  8\ell_3^2    =&&
- \Frac{128 f_\chi^2 F^2}{\overline{M}_V^{\,\,2}}
- 64 \Frac{c_m(2c_d^3-2c_d^2c_m-c_dc_m^2+2c_m^3)  }{ \overline{M}_S^4}
\nonumber \\ &&
\hspace*{+1cm}
- \Frac{32 e^S_m F^2}{\overline{M}_S^{\,\,4}}\,
(c_m^2 - c_d c_m + c_d^2)
 \, ,
\eqa
with the term $ - 128 f_\chi^2 F^2/\overline{M}_V^{\,\,2}$ in the right-hand side
absent in the antisymmetric tensor formalism (n=3).  However, for the case of our
phenomenological lagrangian, its numerical value is absolutely negligible.
Thus, the vector contribution comes mainly from the LEC
${     r_4 = \frac{\lambda_{V\pi\pi}^2 F^2}{M_V^{7-n}}
\left(1+\epsilon_V-(n-5) e^V_m
- \frac{ 2 \overline{M}_V^{\,\,2}}{\overline{M}_S^{\,\,2}}\delta F_{(2)}\right)     }$,
which is extracted from the large--$N_C$ partial wave analysis in a very reliable
way~\cite{matching-op6}.   Likewise, although $r_F$ is obtained by means of the
resonance lagrangian~(\ref{lagr}),  the large errors we found for $\delta F_{(4)}$ in
Eq.~(\ref{dF2+dF4}) make its estimate slightly conservative.
For $\ell_3$ we use,
\bqa \label{lis}
\ell_3=4\frac{c_m (c_m-c_d)}{\overline{M}_S^2}\,,\qquad
\eqa
where we converted  the $SU(3)$ large--$N_C$ estimate from Ref.~\cite{ecker89}
into $SU(2)$ LECs~\cite{SU3-chpt}.
Possible pseudo-scalar resonances have been neglected.
Putting all this information together,
one obtains the value for $r_1$ reported in the text in Eq.~(\ref{eq.r1+split}).

\subsection{Extraction of $r_{S2}$: $\pi\pi$ scalar form factor}
\label{app.SFF}

We proceed now to the calculation of the pion scalar form-factor by means
of the resonance chiral theory lagrangian~(\ref{lagr})~\cite{ecker89,bijri,op6rxt}.
We repeat the calculation in detail, including also the contributions from the
scalar tadpole operator $c_m$~\cite{juanjotadp,bernardtadp}, and obtain
\bear
\mF_S(s)&=& 2 B_0 \left[ 1+ \frac{8 c_m(c_m-c_d)m_\pi^2}{\overline{M}_S^{\,\,2} F^2} +
\Frac{4 c_m}{F_\pi^2} \frac{(c_d s -2 c_d m_\pi^2 + 2c_m
m^2)}{M_\sigma^2-s} \,\,\, +\,\,\, \cO\left(\frac{m_\pi^4}{M_S^4}\right)
\right]\, .
\nn\\
\label{fstadpole}
\eear
Matching the resonance description in
Eq.(\ref{fstadpole}) and the $\chi$PT result for the $\pi\pi$
scalar from-factor~\cite{bijff} leads to the LECs
\be r_{S3}=
4F^2\frac{ c_mc_d }{\overline{M}_S^{\,\,4}},
\ee
\be\label{rs2}
r_{S2} -4\ell_3\ell_4=
\Frac{8c_m (c_m-c_d)F^2}{ \overline{M}_S^4}+\Frac{32
c_m^2c_d(c_d-2c_m)}{\overline{M}_S^{\,\,2}}
\,+\, \Frac{16 c_d c_m F^2}{\overline{M}_S^{\,\, 4}}\, e^S_m\,,
\ee
with $r_{S3}$ consistent with the value from Ref.~\cite{bijff}.
Substituting the large--$N_C$ value of $\ell_3$ from Eq.~(\ref{lis}) and
$\ell_4=4c_m c_d/\overline{M}_S^{\,\,2}$
leads to the prediction for $r_{S2}$  shown in the text in Eq.(\ref{eq.rs2+split}).

%%%
%%%\section{Dispersive determination of the $\cO(p^6)$ couplings  $r_{2...6}$}
%%%
%%%Also the explicit expressions have been given in our previous work, to make the current discussion complete we
%%%present the results below:
%%%\bqa
%%%\label{r2rf}  \hspace*{-1cm} r_2-2r_f &=&  \frac{64
%%%\pi f^6 \overline{\Gamma}_S}{\overline{M}_S^7} \left( 1+
%%%\frac{\beta_{\rm S}}{3} +\frac{\gamma_{\rm S}}{6}   \right)
%%% +\frac{\pi f^6
%%%\overline{\Gamma}_V}{\overline{M}_V^7} \left( 7584 + 1248 \beta_{\rm
%%%V} + 144 \gamma_{\rm V}       \right)\,,
%%%\nonumber \\
%%%r_3&=&\frac{64 \pi f^6 \overline{\Gamma}_S}{3\overline{M}_S^7} \left(
%%%1+\frac{\beta_{\rm S}}{2}\right) \, - \, \frac{768 \pi f^6
%%%\overline{\Gamma}_V}{\overline{M}_V^7} ( 1 + \frac{3  \beta_{\rm
%%%V}}{32} )  \, ,
%%%\nonumber \\
%%% r_4&=&\frac{192 \pi f^6 \overline{\Gamma}_V}{\overline{M}_V^7}
%%% \left( 1 +\frac{\beta_{\rm V}}{8}\right)
%%%\, ,
%%%\nonumber \\
%%%r_5&=&\frac{32 \pi f^6 \overline{\Gamma}_S}{3\overline{M}_S^7}+\frac{36 \pi f^6 \overline{\Gamma}_V}{\overline{M}_V^7}
%%%\, ,
%%%\nonumber \\
%%%r_6&=&\frac{12 \pi f^6 \overline{\Gamma}_V}{\overline{M}_V^7} \,  .
%%%\eqa
%%%

\section{$\cO(p^6)$ low-energy constant running}

The variation of the $r_i^r(\mu)$ with $\mu$ is given by the general pattern~\cite{bijri,bijff}
\be
r_i^r(\mu_1)\, -\, r_i^r(\mu_2) \,\,\ =\,\,\
\Frac{K_i^{(L)}}{ (16\pi^2)^2} \,\ln{\Frac{\mu_1 }{\mu_2 }}
\,\,\,+\,\,\,
\Frac{K_i^{(2)}}{ (16\pi^2)^2}\,\left[\ln^2\Frac{\mu_1 }{m_\pi }\,
-\, \ln^2\Frac{\mu_2 }{m_\pi }\right]\, ,
\ee
with the scale invariants
\bear
K_{1}^{(L)} = \Frac{193}{27} + \Frac{104}{9} \bar{\ell}_1 + \Frac{112}{9} \bar{\ell}_2
-6 \bar{\ell}_3 + 2\bar{\ell}_4\,,&
\qquad \qquad
&K_{1}^{(2)}  =-20\, ,
\nn\\
\nn\\
K_{2}^{(L)} = -\Frac{556}{27} - \Frac{68}{3} \bar{\ell}_1 - \Frac{248}{9} \bar{\ell}_2
+ 7 \bar{\ell}_3 - 2\bar{\ell}_4\,,
&\qquad\qquad&
K_{2}^{(2)} = \Frac{407}{9}\, ,
\nn\\
\nn\\
K_{3}^{(L)} = \Frac{755}{108} + \Frac{100}{9} \bar{\ell}_1 + \Frac{44}{3} \bar{\ell}_2\,,
&\qquad\qquad&
K_{3}^{(2)} = -\Frac{232}{9}\, ,
\nn\\
\nn\\
K_{4}^{(L)} = -\Frac{1}{108} - \Frac{2}{9} \bar{\ell}_1 - \Frac{4}{9} \bar{\ell}_2\,,
&\qquad\qquad&
K_{4}^{(2)} = \Frac{2}{3}\, ,
\nn\\
\nn\\
K_{5}^{(L)} = -\Frac{29}{432} - \Frac{7}{4} \bar{\ell}_1 - \Frac{107}{36} \bar{\ell}_2\,,
&\qquad\qquad&
K_{5}^{(2)} =  \Frac{85}{18}\, ,
\nn\\
\nn\\
K_{6}^{(L)} = -\Frac{79}{432} - \Frac{5}{36} \bar{\ell}_1 - \Frac{25}{36} \bar{\ell}_2\,,
&\qquad\qquad&
K_{6}^{(2)} = \Frac{5}{6}\, ,
\nn\\
\nn\\
K_{S_2}^{(L)} = \Frac{148}{27} + \Frac{62}{9} \bar{\ell}_1 + 2\bar{\ell}_3 + 2\bar{\ell}_4\,,
&\qquad\qquad&
K_{S_2}^{(2)} = -\Frac{166}{9}\, .
\eear

\section{Chiral corrections to the   decay width}
\label{lambdaV8}

In the article, the splitting in the vector decay widths due to quark mass effects
was parametrized at NLO by one single parameter $\epsilon_V$
(or equivalently $f_\chi$).  This was true for the Proca resonance lagrangian
in Eq.~(\ref{lagr}).  However, in the case of a more general hadronic action,
one single parameter $\epsilon_V$ does not seem  enough to describe
the NLO chiral corrections to the decay widths.
In the Proca field realization, in addition to the operator
$ f_\chi\bra \hat{V}_\mu [ u^\mu, \chi_- ] \ket$, there could exist
higher derivative operators  such as
$\bra \hat{V}_{\mu\nu}\{ \chi_+,  u^\mu  u^\nu \} \ket$, which
also contribute to the chiral corrections to the decay widths at NLO in  $m_\pi^2$.
Nonetheless, we will focus our digression about higher terms in the lagrangian
in the antisymmetric tensor formalism,
where the  possible operators related to the NLO quark mass corrections  to the
vector width have been already constructed in Ref.~\cite{op6rxt}:
\bqa
i\lambda_8^V \bra V_{\mu\nu} \{ \chi_+ , u^\mu u^\nu\}\ket +
i \lambda_9^V \bra V_{\mu\nu} u^\mu\chi_+u^\nu \ket
+  \lambda_{10}^V \bra V_{\mu\nu}[ u^\mu, \nabla^\nu \chi_- ]\ket\,.
\eqa

However, not all  these three couplings are observable in the
partial decay widths we analyzed in the text, which now become
%%%\bqa
%%%\Gamma_{\rho\to\pi\pi} \,&=&\,
%%%\Frac{ {G_V^{\rm eff}}^2 M_\rho^3 \rho_{\rho\pi\pi}}{48\pi F_\pi^4}  \,
%%%\left[1\, +\,\Frac{4\sqrt{2} \,\widetilde{\lambda}_8^V m_\pi^2}{G_V^{\rm eff}}\,\right]^2\, ,
%%%\nonumber \\
%%%\Gamma_{K^*\to K\pi} \,&=&\,
%%%\Frac{ {G_V^{\rm eff}}^2 M_{K^*}^3 \rho_{K^*K\pi}}{64\pi F_K^2 F_\pi^2}  \,
%%%\left[1\, +\,\Frac{2\sqrt{2} \, \widetilde{\lambda}_8^V (m_\pi^2 + m_K^2)}{G_V^{\rm eff}}\, \right]^2\, ,
%%%\nonumber \\
%%%\Gamma_{\phi\to K \overline{K}} \,&=&\,
%%%\Frac{ {G_V^{\rm eff}}^2 M_{\phi}^3 \rho_{\phi K\overline{K}}}{48\pi F_K^4}  \,
%%%\left[1\, +\,\Frac{4\sqrt{2} \,\widetilde{\lambda}_8^V m_K^2 }{G_V^{\rm eff}}\, \right]^2\, ,
%%%\eqa
\bqa
\Gamma_{V\to\phi_1\phi_2} &=& C_{V 1 2}\,
\Frac{ M_V^{\,\, 3} \, \rho_{\rm V 12}^3 }{48\, \pi \,  F_{1}^2 \, F_{2}^2}
\,\,\, {G_{V}^{\rm eff}}^2\,\,\, \left[1 +
\Frac{2\sqrt{2} \, (2\lambda_8^V+\lambda_{10}^V)}{G_V^{\rm eff}}\,\,  (m_1^2+m_2^2)
\,+  \cO(m_P^4)\right]^2\,,
\nn\\
\eqa
with the corresponding Clebsch-Gordan $C_{V 1 2}$ for the channels under consideration,
$C_{\rho\pi\pi}=1$, $C_{{K^*} K\pi}=3/4$ and $C_{\phi K\overline{K}}=1$.
Instead of the $m_\pi$--independent coupling $G_V$, which we had before in
Eq.~(\ref{avdef}), one has now the effective combination
\be
G_V^{\rm eff}  \, =\,  G_V
\, +\,  2\sqrt2 \, ( \lambda_9^V- 2 \lambda_8^V)\,  m_\pi^2\,.
\ee
%%provide by the combinations
%%\be
%%\widetilde{\lambda}_8^V= 2\lambda_8^V+\lambda_{10}^V\,,\qquad
%%\widetilde{\lambda}_9^V= \lambda_9^V+\lambda_{10}^V\,.
%%\ee
The combination of resonance couplings $(2\lambda_8^V+\lambda_{10}^V)$ rules
the splitting of our studied decays and determines the chiral symmetry
breaking parameter
${  f_\chi^{\rm n=3}= (2\lambda_8^V+\lambda_{10}^V) \overline{M}_V^{\,\,2}    }$.

Hence,  the  $\rho \to \pi\pi$, $K^*\to K\pi$ and $\phi\to K \overline{K}$
partial decay widths only allow the determination of $G_V^{\rm eff}$,
not its chiral limit value $G_V$.  A similar situation would happen
with  the scalar sector and the Proca field description for vectors,
where the presence of higher operators could make us observe
in our analysis some effective combinations
$c_d^{\rm eff}$ and $g_V^{\rm eff}$, instead of the quark mass independent couplings
$c_d$ and $g_V$. In any case, the difference is assumed to be small in the
present work and its further study  remains beyond the scope of this article.


\begin{thebibliography}{26}
\bibitem{matching}
    Z.~H.~Guo, J.~J.~Sanz Cillero and H.~Q.~Zheng,
    {\it Partial-waves and large--$N_C$ resonance sum rules},
    JHEP {\bf 0706} (2007) 030.

%%%\bibitem{sigma}
%%%    Z. H. Guo,
%%%    {\it Is the sigma meson dynamically generated?},
%%%    AIP. Conf. Proc. {\bf 1030} (2008) 408.

\bibitem{matching-op6}
    Z.~H.~Guo, J.~J.~Sanz Cillero and H.~Q.~Zheng,
    {\it $\cO(p^6)$ extension of the large--$N_C$  partial-wave dispersion relations}\,,
    Phys. Lett. {\bf B 661} (2008) 342.




\bibitem{bijri}
    J.~Bijnens, G.~Colangelo, G.~Ecker, J.~Gasser and M.~E.~Sainio,
    {\it  Pion pion scattering at low-energy},
    Nucl.Phys.{\bf B 508} (1997) 263, Erratum-ibid.{\bf B 517}(1998) 639.


\bibitem{bijff}
    J.~Bijnens, G.~Colangelo and P.~Talavera,
    {\it The Vector and scalar form-factors of the pion to two loops},
    JHEP {\bf 05} (1998) 014.



\bibitem{NC}
  G.~'t Hooft,
  {\it A planar diagram theory for strong interactions},
  Nucl.\ Phys.\ {\bf B 72}, 461 (1974); {\bf 75}, 461 (1974);
  \\
    G. 't Hooft,
    {\it A Two-Dimensional Model for Mesons},
     Nucl. \ Phys. \   {\bf B 75} (1974) 461;
  \\
  E.~Witten,
  {\it Baryons in the $1/N$ expansion},
  Nucl.\ Phys.\  {\bf B 160}, 57 (1979).


\bibitem{slgasser}
    G.~Colangelo, J.~Gasser and H.~Leutwyler,
    {\it $\pi\pi$ scattering},
    Nucl. Phys.{\bf B 603} (2001) 125.



\bibitem{sigmaleut}
    I. Caprini, G. Colangelo and H. Leutwyler,
    {\it Mass and width of the lowest resonance in QCD},
    Phys. Rev. Lett. {\bf 96} (2006) 132001.

\bibitem{sigmazheng}
    Z.~G.~Xiao and H.~Q.~Zheng,
    {\it Left-hand singularities, hadron form-factors
    and the properties of the sigma meson},
    Nucl.Phys.  {\bf A 695} (2001) 273;
\\
    Z.~Y.~Zhou {\it et al.},
    {\it  The Pole structure of the unitary, crossing
    symmetric low energy $\pi\pi$ scattering amplitudes},
    JHEP {\bf 02} (2005) 043.



\bibitem{pdg}
    C. Amsler {\it et al.} (Particle Data Group),
    {\it Review of Particle Properties},
    Phys. Lett. {\bf B 667} (2008) 1.


\bibitem{ecker89}
    G. Ecker {\it et al.},
    {\it The Role of Resonances in Chiral Perturbation Theory},
    Nucl. Phys. {\bf B 321} (1989)311.

\bibitem{op6rxt}
    V. Cirigliano {\it et al.},
    {\it Towards a consistent estimate of the chiral low-energy constants},
    Nucl. Phys. {\bf B 753} (2006) 139.



\bibitem{bernardtadp}
    V. Bernard, N. Kaiser and Ulf G. Meissner,
    {\it Chiral perturbation theory in the presence of
    resonances: Application to $\pi\pi$ and $\pi K$ scattering},
    Nucl. Phys. {\bf B 364} (1991) 283.

\bibitem{juanjotadp}
    J.J. Sanz-Cillero,
    {\it Pion and kaon decay constants: Lattice versus resonance chiral theory},
    Phys. Rev. {\bf D 70} (2004) 094033.


\bibitem{spin1fields}
  G.~Ecker, J.~Gasser, H.~Leutwyler, A.~Pich and E.~de Rafael,
  {\it Chiral Lagrangians for massive spin 1 fields},
  Phys.\ Lett.\   {\bf B 223}, 425 (1989).


\bibitem{mass-split}
    V. Cirigliano, G. Ecker, H. Neufeld and A. Pich,
    {\it  Meson resonances, large--$N_C$ and chiral symmetry},
    JHEP {\bf 0306} (2003) 012.

\bibitem{cdcm}
    M. Jamin, J.A. Oller and A. Pich,
    {\it Strangeness changing scalar form-factors},
    Nucl.~Phys.~{\bf B 622} (2002) 279.

\bibitem{cdcm2}
    M.~Jamin, J.A.~Oller and A.~Pich,
    {\it S wave $K\pi$ scattering in chiral perturbation theory with resonances},
    Nucl.\ Phys.\  {\bf B 587} (2000) 331-362.

\bibitem{ND}
    J.A. Oller and  E. Oset,
    {\it N/D description of two meson amplitudes and chiral symmetry},
    Phys.\ Rev.\ {\bf  D  60} (1999) 074023.
%    [arXiv:hep-ph/9809337].

\bibitem{cd-Kaiser}
    R.~Kaiser,
    {\it Large $N_C$ in chiral resonance Lagrangians},
    {\bf Trento 2004, Large--$N_C$ QCD} 144-159.

\bibitem{Sdecays}
    S. Ivashyn and A.Yu. Korchin,
    {\it Radiative decays with light scalar mesons and singlet-octet
    mixing in ChPT},
    Eur.\ Phys.\ J.\  {\bf C 54} (2008) 89-106.


\bibitem{fetaleut}
    H. Leutwyler,
    {\it  On the 1/N expansion in chiral perturbation theory},
    Nucl.~Phys.~Proc.~Suppl.~{\bf 64} (1998) 223;
\\
    R. Kaiser and H. Leutwyler,
    {\it Pseudoscalar decay constants at large--$N_C$},
    {\bf Adelaide 1998, Nonperturbative methods in quantum field theory}  15-29
    [arXiv:hep-ph/9806336].


\bibitem{thetaeta}
  F.G Cao and A.I. Signal,
  {\it Two analytical constraints on the $\eta$--$\eta'$ mixing},
  Phys. Rev. {\bf D 60} (1999) 114012.


\bibitem{thetaeta2}
%%    R.~Escribano,
%%    {\it $\eta$--$\eta'$ mixing from $V\to P\gamma$ and $J/\psi\to VP$ decays},
%%    [arXiv:0812.0628~[hep-ph]].
%%
    R. Escribano,
    {\it $J/\psi\to VP$ decays and the quark and gluon content of the eta and eta-prime},
    [arXiv:0807.4201 [hep-ph]].





\bibitem{PI:02}
  A. Pich,
  {\it Colorless mesons in a polychromatic world},
  {\bf Tempe 2002, Phenomenology of large--$N_C$ QCD} 239-258
 [arXiv:hep-ph/0205030].



\bibitem{SPP}
    V. Cirigliano, G. Ecker, M. Eidemuller, R. Kaiser, A. Pich and J. Portoles,
    {\it The $\bra SPP\ket$ Green function and SU(3) breaking in $K_{\ell3}$ decays},
    JHEP {\bf 0504} (2005) 006.


\bibitem{gasser84}
    J.~Gasser and H.~Leutwyler,
    {\it Chiral Perturbation Theory to One Loop},
    Annals~Phys.~{\bf 158} (1984) 142.



\bibitem{Cata+Peris}
    O. Cata and S. Peris,
    {\it An Example of resonance saturation at one loop},
    Phys. Rev. D {\bf 65} (2002) 056014.

\bibitem{L8}
    I. Rosell, J.J. Sanz-Cillero and A. Pich,
    {\it Towards a determination of the chiral couplings at NLO in $1/N_C$:
    $L^r_8(\mu)$ and $C^r_{38}(\mu)$},
    JHEP {\bf 0701} (2007) 039;
    \\
    J.J. Sanz-Cillero,
    {\it Resonance form-factors: $L_8^r(\mu)$ determination at next-to-leading order
    in $1/N_C$},
    [arXiv:hep-ph/0610304].

\bibitem{L9}
    I. Rosell, J.J. Sanz-Cillero and A. Pich,
    {\it Quantum loops in the resonance chiral theory: The Vector form-factor},
    JHEP {\bf 0408} (2004) 042.



\bibitem{L10}
    I. Rosell, J.J. Sanz-Cillero and  A. Pich,
    {\it Form-factors and current correlators: Chiral couplings $L^r_{10}(\mu)$
    and $C_{87}^r(\mu)$ at NLO in $1/N_C$},
    JHEP {\bf 0807} (2008)  014.


\bibitem{Kampf-L2}
    K. Kampf and  B. Moussallam,
    {\it Tests of the naturalness of the coupling constants in ChPT at $\cO(p^6)$},
    Eur. Phys. J. C {\bf 47} (2006) 723-736.


\bibitem{op6-reno}
    J. Bijnens, G. Colangelo and G. Ecker,
    {\it Renormalization of chiral perturbation theory to order $ p^6 $},
    Annals Phys. {\bf 280} (2000) 100-139.


\bibitem{SU3-chpt}
    J.~Gasser and H.~Leutwyler,
    {\it Chiral Perturbation Theory: Expansions in the Mass of the Strange Quark},
    Nucl. Phys. {\bf B  250} (1985) 465.






\end{thebibliography}
\end{document}